# Improving the State of the Art for Training Human-AI Teams

**Technical Report #3**

# Analysis of Testbed Alternatives

**June 2023**

*Lillian Asiala[*], Jim McCarthy[*], & Lixiao Huang[^]*

[*]*Sonalysts, Inc.*

[^]*Arizona State University*

This Page Intentionally Blank

# TABLE OF CONTENTS

# LIST OF FIGURES

# LIST OF TABLES

This Page Intentionally Blank

# ACRONYMS

# ACKNOWLEDGEMENTS

We would like to acknowledge the individuals who donated their time and expertise to help us identify and assess the various testbeds discussed in this report:

                        Sarah Bibyk

                        Katherine Cox

                        David Grimm

                        Lixiao Huang

                        Joan Johnston

                        Andrea Krausman

                        Marc Steinberg

                        Melissa Walwanis

We would also like to express our appreciation to those participants who asked to remain anonymous.

This Page Intentionally Blank

# Abstract

Sonalysts is working on an initiative to expand our current expertise in teaming to Human-Artificial Intelligence (AI) teams by developing original research in this area. To provide a foundation for that research, Sonalysts is investigating the development of a Synthetic Task Environment (STE). In a previous report, we documented the findings of a recent outreach effort in which we asked military Subject Matter Experts (SMEs) and other researchers in the Human-AI teaming domain to identify the qualities that they most valued in a testbed. A surprising finding from that outreach was that several respondents recommended that our team look into existing human-AI teaming testbeds, rather than creating something new. Based on that recommendation, we conducted a systematic investigation of the associated landscape. In this report, we describe the results of that investigation. Building on the survey results, we developed testbed evaluation criteria, identified potential testbeds, and conducted qualitative and quantitative evaluations of candidate testbeds. The evaluation process led to five candidate testbeds for the research team to consider. In the coming months, we will assess the viability of the various alternatives and begin to execute our program of research.

This Page Intentionally Blank

## 1 BACKGROUND

A consensus report produced for the Air Force Research Laboratory (AFRL) by the National Academies of Sciences, Engineering, and Mathematics documented a prevalent and increasing desire to support Human-AI teaming across military service branches (NASEM, 2021). Sonalysts has begun an internal initiative to explore ways to maximize the performance of Human-AI teams. To provide a foundation for our research, Sonalysts is exploring options for a Synthetic Task Environment (STE) that can serve as a testbed.

Building on the results of recent surveys of Subject-Matter Experts (SMEs; McCarthy & Asiala, 2023a) and researchers within the Human-AI teaming domain (McCarthy & Asiala, 2023b), in this report we explore the testbed qualities on which we will focus, identify potential STEs, and conduct a quantitative comparison of the candidate STEs.

## 2 QUALITIES OF A DESIRABLE TESTBED

In this section we outline a range of features that researchers may wish to consider when selecting a STE, and identify the features on which we will focus within our initiative. We will begin by introducing a notional taxonomy that researchers can use to organize various testbeds, and then explore specific STE features that are likely to be desirable. We conclude with a brief discussion of the STE dimensions and features on which we will focus in support of the initiative.

### 2.1 Testbed Taxonomy

Discussions with various researchers in the field (*e.g.,* M. Steinberg, 2023) led us to create the STE taxonomy shown in Figure 1. In this conception of the taxonomy, the dimensions are the level of interdependency among team members (humans and agents), the relevancy of the task presented within the testbed, and the sophistication of the agents that populate the testbed.

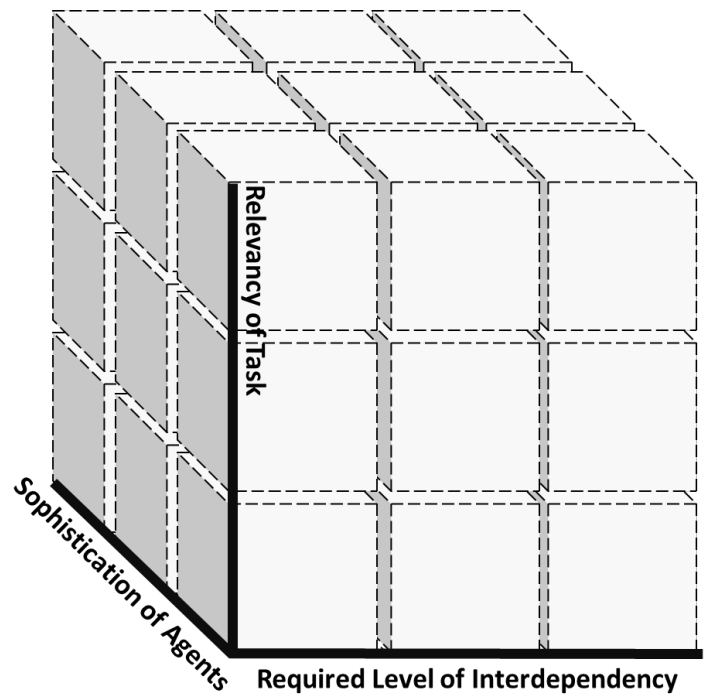

**Figure 1: Notional STE Taxonomy**

Of these, the ***level of interdependency dimension*** is likely the most important. The interdependency dimension reflects the extent to which team members are dependent on others to complete the collective task. Fortunately, earlier work allows us to operationalize this dimension. First, consider the work of Saavedra, Early, and Van Dyne (1993). Working in the business-focused "group work" domain, these researchers, citing the earlier work of Thompson (1967) and Van de Ven et al. (1976), discuss the range of interdependence shown in Figure 2. At one extreme (*pooled workflow*), there is little to no interdependence; each individual completes the assigned task independently and the "team product" is the sum of the individual effort (*e.g.,* the individual workers in a given call center). In the *sequential workflow*, individuals are responsible for producing their work product and passing it to the next team member for further processing (*i.e.,* a classic "assembly line" process). In this configuration, there is a one-way flow of information, resources, work product, *etc*. A greater level of interdependence is present in the remaining two cases. In the *reciprocal model*, there is a two-way flow between "adjacent" workers. This "give and take" requires greater levels of coordination and cooperation than the one-way flow that is present in the sequential model. Finally, in the *team model* there is a multi-way sharing of information, resources, work product, *etc.* Each member of the team can participate in a give and take with any other member of the team (see also, Singh, Sonenber, & Miller, 2017).

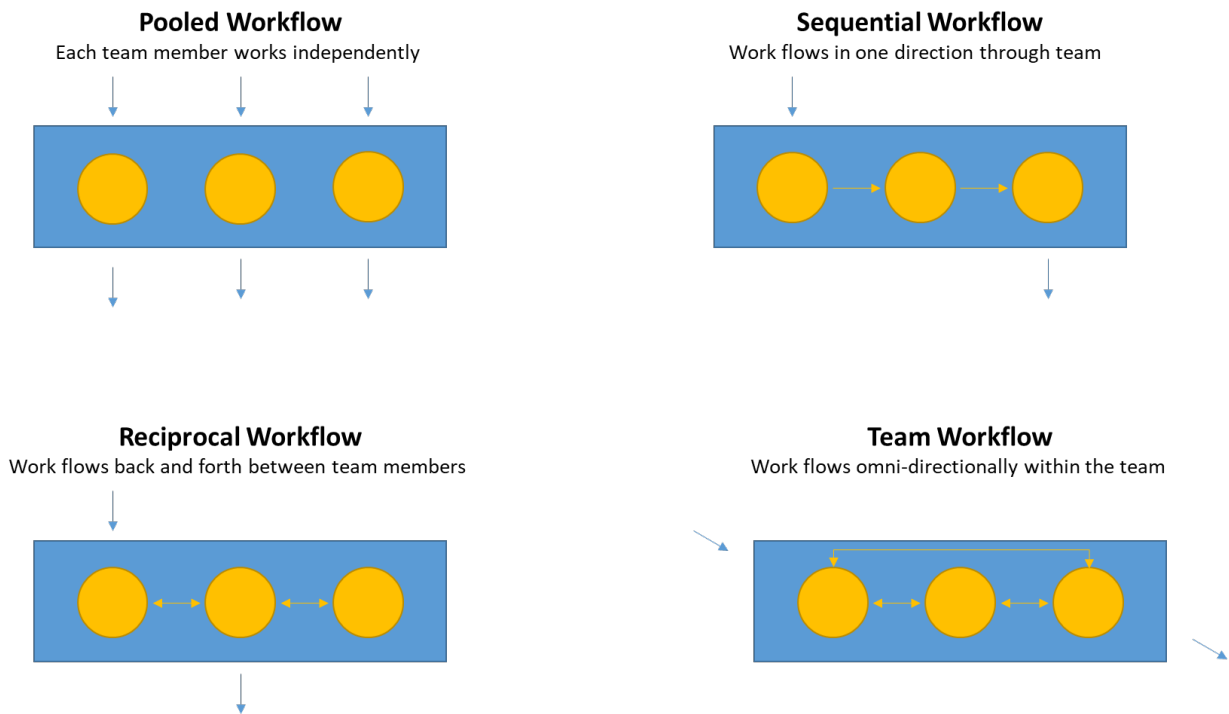

**Figure 2: Workflow Hierarchy as Defined by Thompson (1967) and Van de Ven et al. (1976) - adapted from Saavedra, Early, and Van Dyne (1993)**

Working in a military setting some years later, Arthur et al. (2005) defined two factors that influence interdependency. The first factor ("workflow") was adapted from the work of Saavedra, Early, and Van Dyne (1993); Thompson (1967); Van de Ven et al. (1976); and Tesluk, Mathieu, and Zaccaro (1997). To operationalize the concept, Arthur et al. (2005) created a self-report measure. Using the provided chart (see Figure 3), team participants indicated the workflow pattern that most closely resembled what occurred in their team. As measured with this instrument, workflow reflects the team members' *perceptions* of how work must flow to complete a given task. As such, it is not a measure of the optimal workflow nor the actual workflow.

| Team Workflow Pattern | Description | Illustration | RESPONSE |
|---|---|---|---|
| 1. Not a Team Task/Activity. | Work and activities are **NOT** performed as a member of a team; they are performed alone outside the context of the team. Work and activities are performed by an individual working **ALONE**, **NOT** in a team. | Work Received by Individual ↓ ○ ↓ Work Leaves Individual | ① |
| 2. Pooled/Additive Interdependence. | Work and activities are performed separately by all team members and work does not flow between members of the team. | Work Enters Team ↓↓↓↓ [○ ○ ○ ○] ↓↓↓↓ Work Leaves Team | ② |
| 3. Sequential Interdependence. | Work and activities flow from one member to another in the team, but mostly in one direction. | Work Enters Team ↓ [○→○→○→○] ↓ Work Leaves Team | ③ |
| 4. Reciprocal Interdependence. | Work and activities flow between team members in a back-and-forth manner over a period of time. | Work Enters Team ↓↓↓↓ [○⇌○⇌○⇌○] ↓ Work Leaves Team | ④ |
| 5. Intensive Interdependence. | Work and activities come into the team and members must diagnose, problem solve, and/or collaborate as a team in order to accomplish the team's task. | Work Enters Team ↓ [○ ○ / ○ ○] ↓ Work Leaves Team | ⑤ |

**Figure 3: Workflow Measure (Adapted from Arthur et al., 2005)**

To complement the workflow measure, Arthur et al. (2005) developed a measure of "Team Relatedness." Figure 4 is the relatedness measure. The relatedness scale asked respondents to assess the degree to which a given task can be performed by an individual. Within this measure, five levels of "relatedness" are defined for raters. The raters must then assign one of the five values to each task that must be performed.

For each task/activity presented below, please shade the number corresponding to the:
(a) IMPORTANCE of the task/activity to the performance of your job, and
(b) TEAM RELATEDNESS of the task/activity = The extent to which successful team performance requires you to work with members of the team in order to optimally perform the specified task.

IMPORTANCE
① = Not at all important
② = Of little importance
③ = Somewhat important
④ = Very important
⑤ = Of highest importance

TEAM RELATEDNESS
① = Not required to work with team members at all for optimal performance
② = Required to work with team members very little for optimal performance
③ = Somewhat required to work with team members for optimal performance
④ = Required to work with team members quite a bit for optimal performance
⑤ = Very much required to work with team members for optimal performance

| Tasks/Activities | Importance | Team Relatedness |
|---|---|---|
| 1. Aiming | ① ② ③ ④ ⑤ | ① ② ③ ④ ⑤ |
| 2. Lasing | ① ② ③ ④ ⑤ | ① ② ③ ④ ⑤ |
| 3. Buttoning the tank | ① ② ③ ④ ⑤ | ① ② ③ ④ ⑤ |
| 4. Visually locating targets | ① ② ③ ④ ⑤ | ① ② ③ ④ ⑤ |
| 5. Timing tank movement with the gunner's shots | ① ② ③ ④ ⑤ | ① ② ③ ④ ⑤ |
| 6. Entering battle position | ① ② ③ ④ ⑤ | ① ② ③ ④ ⑤ |
| 7. Emitting smoke screen | ① ② ③ ④ ⑤ | ① ② ③ ④ ⑤ |
| 8. Slewing the turret | ① ② ③ ④ ⑤ | ① ② ③ ④ ⑤ |
| 9. Destroying enemy tanks | ① ② ③ ④ ⑤ | ① ② ③ ④ ⑤ |
| 10. Completing the mission | ① ② ③ ④ ⑤ | ① ② ③ ④ ⑤ |

**Figure 4: Relatedness Measure (Adapted from Arthur et al., 2005)**

Both instruments produced high levels of interrater reliability (0.80 – 0.96). Critically, tasks that the research team designed to have a greater level of interdependence were rated higher on both scales than those that were designed to more accurately represent individual taskwork. Interestingly, the two instruments were only moderately correlated with each other, indicating that they assess different aspects of the task.

The *sophistication dimension* refers to the sophistication of the agents that are operational within the testbed. It is a factor in the taxonomy to the extent that the testbed limits which agents researchers can integrate or the capabilities that those agents can bring to bear. This appears to be a largely qualitative dimension; in our research, we have not discovered robust operational definitions of sophistication.

The *relevancy dimension* refers to how directly the testbed task transfers to the application environment. Again, while a given research environment is likely to be more or less closely related to a given transfer environment, this distinction is largely qualitative and not operationalized.

Returning our focus to Figure 1, a trivial task environment in which all participants can complete their tasks independently and the agents are unsophisticated would represent the lower left portion of the taxonomy cube (see Figure 5). On the other hand, a sophisticated task environment that 1) faithfully recreates a notional Joint All-Domain Command and Control (JADC2) operational environment, 2) requires participants to solve realistically complex challenges, and 3) allows for the incorporation of sophisticated agents that have both "taskwork" and "teamwork" skills would be represented by the green box in the upper right corner of the taxonomy (see Figure 6).

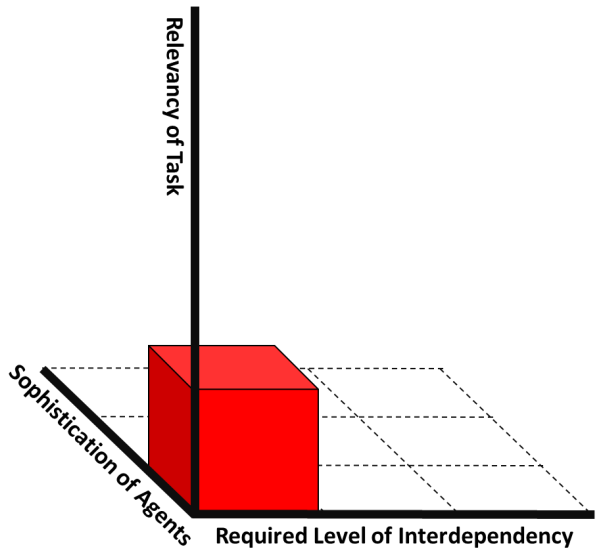

Figure 5: Testbed Low on All Dimensions

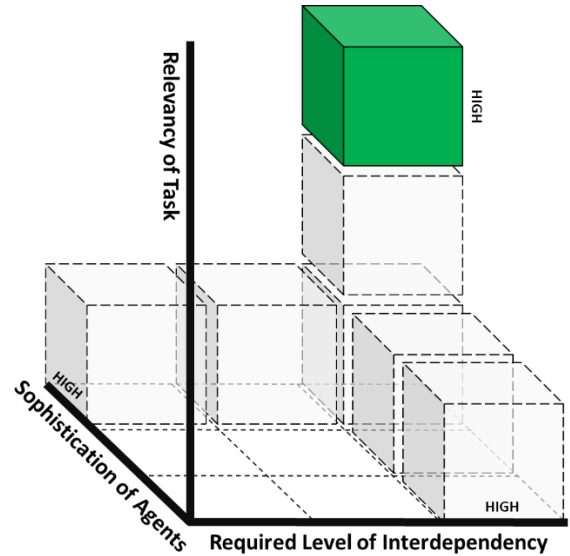

Figure 6: Testbed High on All Dimensions

With this taxonomy in mind, it becomes clear that different regions of the taxonomy might be aligned with different research goals. For example, more **basic research** on the features of agents that lead them to be perceived as trusted teammates may require high levels of agent sophistication and interdependency, but only moderate levels of relevancy (the "back right" corner of the middle slice of the cube; see Figure 7). However, if we are conducting **very applied research** to evaluate a specific technology for a specific application, we may want to be high on all three dimensions (the cell at the far upper right of the cube; see Figure 8).

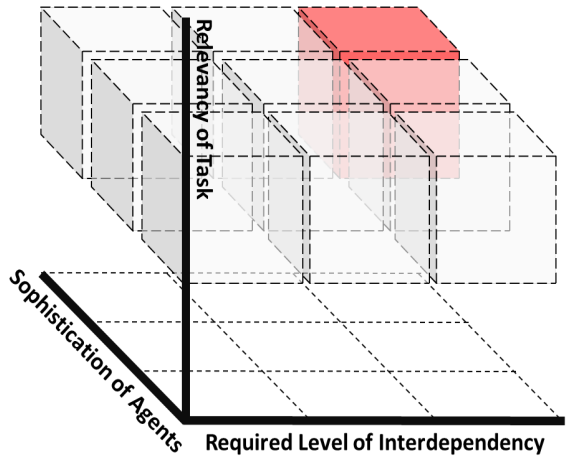
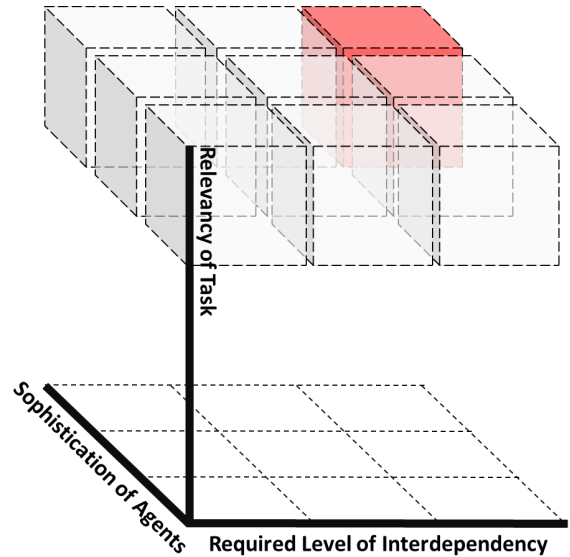

Figure 7: More Basic Research    Figure 8: More Applied Research

We will return to a consideration of the portions of the taxonomy that are relevant to our initiative in Section 4.

## 2.2  Desirable Features

Prior to developing the SME and Researcher surveys, Sonalysts conducted a fairly broad literature review. The review, coupled with our experience in a range of modeling and simulation efforts, allowed us to predict a range of task domain features that we thought might be useful to include. Within the survey, we asked respondents to indicate the degree to which they agreed or disagreed with our hypotheses about which specific features would be important within the STE. We complemented the Likert-like items with open-ended questions that allowed survey respondents to offer suggestions for additional features that might be useful within the STE.

In support of this effort, the research team revisited the survey results to identify features that we could use to characterize and differentiate candidate environments. We began with the researcher survey responses as a basis for testbed feature identification. During an initial analysis of survey responses, we conducted a thematic analysis of responses to open-ended questions. Our goal was to identify recurring ideas across answers, regardless of the terminology or phrasing that respondents used. We combined the thematic analyses into a consensus list of themes for each question, and the research team mapped individual responses to entries on that list. To create a master list of STE features, we combined the items on the consensus list with the hypothesized features that were presented as Likert-style items eliciting agreement ratings. This analysis yielded 289 individual entries.

We then sorted candidate features into superordinate feature descriptions. We did this so that we could identify and account for overlap between the features mentioned in responses to the open-ended questions and the features represented in Likert-style questions. For example, one Likert-style question asked participants to rate their agreement with the statement "The STE should be implemented as an 'open source' tool." In addition to general agreement with this item, six respondents mentioned that the STE should have an "open source architecture" in their response to other questions in the survey. We categorized these entries within a superordinate feature category "The STE should be 'open source'."

The superordinate feature-categorization process yielded more than one hundred superordinate categories. To create a more reasonable feature set, we dropped categories with less than three combined open-ended responses and Likert-style items, and combined a few categories with a high degree of conceptual overlap. This narrowed the list to 23 features. Since 23 is still an unwieldy number of individual criteria to use to evaluate and compare testbeds, we further consolidated the superordinate categories into the eight broad criteria listed in Table 1. The bullets associated with each criterion reflect the associated finer grained features. We describe the contribution of these criteria to a quantitative testbed evaluation tool in Section 4.

**Table 1: Important STE Characteristics**

| Implementation Factors | System Architecture |
|---|---|
| <ul><li>Speed of Implementation[1]</li><li>Implementation Cost[1]</li></ul> | <ul><li>Open source</li><li>Modular/Flexible</li><li>Permits collaboration across labs</li></ul> |
| Data Collection/Performance Measurement | Task Features |
| <ul><li>Collect outcome measures</li><li>Collect process measures</li><li>Can be instrumented to collect a variety of data, including individual and team status</li></ul> | <ul><li>Task should be easy to learn but potentially challenging</li><li>Task demands should replicate or impose information processing demands analogous to real-world tasks</li><li>Tasks should have adjustable levels of difficulty, uncertainty, time pressure, *etc*.</li><li>Tasks should require significant coordination/interdependency</li><li>Tasks should impose conflicts among individual goals and team goals[1]</li><li>Tasks should require communication among teammates (*e.g.,* spoken, text chat)</li><li>Tasks should include adjustable levels of information density, signal-to-noise ratio, and data reliability[1]</li></ul> |
| Data Processing | Scenario Authoring |
| <ul><li>Facilitate data export</li><li>Include "quick look" visualizations of individual variables and variable pairs</li></ul> | <ul><li>Facilitate the definition of linkages among competencies, scenario events, and assessment routines</li><li>Ease the creation of unique scenarios aligned with research objectives</li></ul> |

---

[1] Indicates items that were not directly assessed/mentioned within the SME and Researcher surveys, but that are likely to be important to our initiative

| Teaming Factors | Agents |
|---|---|
| • Support teams with multiple defined roles (nominally 4-12)<br>• Allow each role to be filled by a human or agent<br>• Distinguish between the source and intended recipient for each communication act | • Facilitate integration of agents with differing capabilities |

It is interesting to note that this list of features has many similarities to others that researchers have produced. See, for example, the work of Cavanah et al. (2020).

In selecting a testbed for Sonalysts' program of Human-AI teaming research, we can consider both the taxonomy introduced in Figure 1 and the important STE characteristics identified in Table 1.

Within the taxonomy, we desire a high level of interdependency. This will ensure that our tasks require the type of teamwork that is most representative of military settings and likely most difficult for agents to smoothly support. Similarly, we will want our testbed to support agents with at least moderate levels of sophistication. As we work to determine which features contribute to Human-AI team performance, we do not want the testbed to constrain the level of sophistication present in the agents. Relevancy will be less important to our research. Figure 9 illustrates a portion of the taxonomy that is most likely to be relevant to our research.

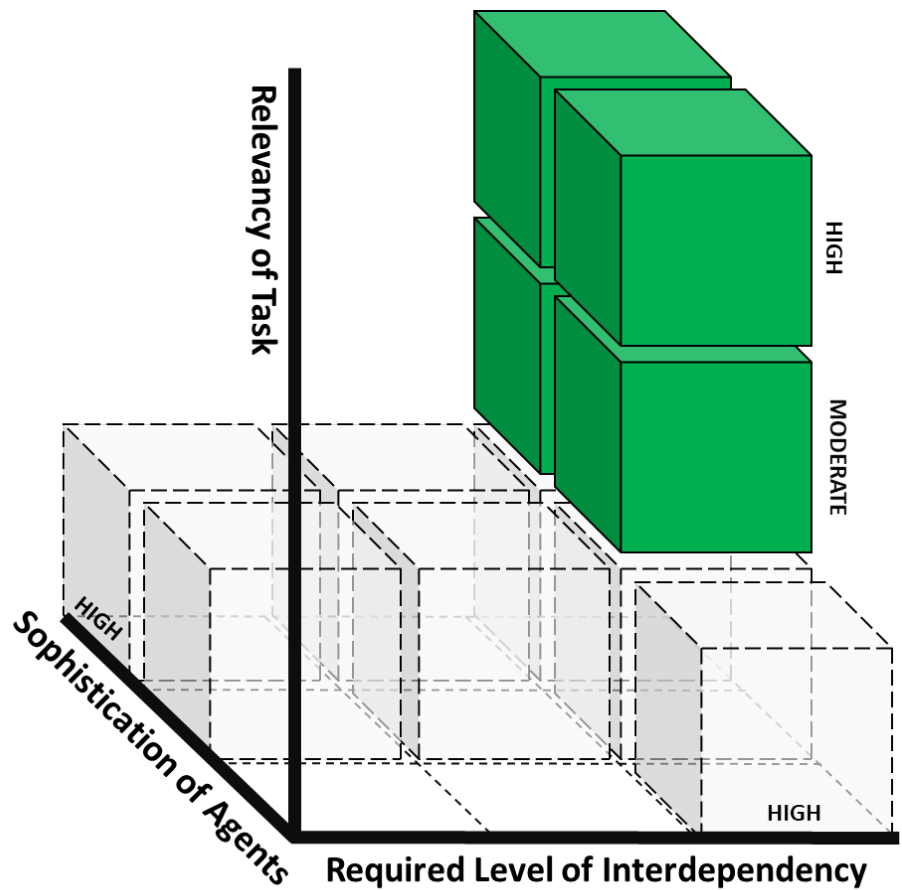

**Figure 9: Targeted Portion of Testbed Taxonomy**

Table 2 summarizes the characteristics on which we want to focus during our search for a relevant testbed. It provides a list of the characteristics we will assess, and lists any particularly important factors when considering each characteristic. We will discuss these characteristics in greater detail in Section 4.

Table 2: Driving Factors in the Selection of a Human-AI Teaming Testbed for Sonalysts' Research Initiative.

| Key Feature | Discussion |
|---|---|
| Data Collection & Performance Measurement | The testbed must perform and store outcome and process assessments on individuals and teams. |
| Implementation Factors | As an internally funded effort, we will want to ensure that our budget is directed to data collection and analysis to the greatest degree possible, rather than siphoned away to the development of tools that may not be widely employed. |
| Teaming Factors | It will be important to have teams comprising at least four members, and we will want a great deal of flexibility in assigning specific roles to humans or agents. We also want to be able to assess communication among team members effectively. |
| Task Features | It will be important for the testbed to provide tasks with a high degree of psychological fidelity and the ability to modulate a number of task factors (*e.g.*, difficulty, uncertainty, time pressure, information density, signal-to-noise ratio, and data reliability). |
| Scenario Authoring | Our team is skilled in the development of scenario-based environments, and the authoring tools can be fairly rudimentary. The STE should support our ability to create scenarios that meet our research objectives, and make connections between competencies, scenario events, and measures of performance. |
| Data Processing | As an internal effort, the ease with which data are exported is not a primary consideration. The team will have the sophistication necessary to pull data from a variety of databases, regardless of type or structure. Further, we can use external tools for data analysis and visualization. |
| System Architecture | A variety of system architectures and approaches are likely to be acceptable, as long as it is not cumbersome to make changes to the testbed as our research progresses. |
| Agents | In keeping with our taxonomy discussion (see Section 2.1), we want to make it relatively easy to "plug-in" agents with a variety of capabilities. |

## 3  QUALITATIVE DESCRIPTION OF POSSIBLE TESTBEDS

With these characteristics in mind, the team began to consider various possible testbeds. We are particularly interested in identifying existing testbeds that may be suitable for our research. Employing these testbeds may reduce development costs and accelerate timelines while offering a literature that would position our findings in a broader context. In this section, we offer narrative descriptions of several such systems (listed alphabetically) and identify those that warrant further scrutiny.

A review of the literature has identified three classes of such testbeds:

1. Testbeds that were specifically developed to support research on Human-AI teaming.
2. Testbeds that were built on a foundation provided by commercial games.
3. Testbeds that were built on a foundation provided by open source games.

We will review each of these categories in turn.

### 3.1 Custom-Developed Testbeds

Researchers specifically developed the first class of testbeds that we wish to consider to assess team performance, especially in the context of Human-AI teaming. These environments are of significant interest because they are likely to include the instrumentation needed for robust data collection.

We considered six testbeds within this category:

1. Agile-Teams Testbed (AT2B)
2. Blocks World for Teams (BW4T)
3. MazeWorld
4. Multi-Agent Computing Environment (MACE)
5. Michigan Intelligent Coordination Experiment (MICE)
6. Reconfigurable Testbed for Teams (RT4T)

We describe each of these in the sections that follow and identify those that have particular potential for use as testbeds within our research.

#### 3.1.1 Agile-Teams Testbed (AT2B)

This system was developed by Cubic's team as part of the Defense Advanced Research Projects Agency's (DARPA) Agile Teams Project (https://www.darpa.mil/program/agile-teams). The effort was not successful and was eventually dropped. Several lessons learned, however, might inform the current effort (Magaha & Macron, 2021):

- Reserve Officer Training Corps (ROTC) instructors who participated in a pilot test of the testbed found that the After-Action Reviews (AARs) that followed game sessions were useful.
- Pilot test participants appreciated some "quick look" visualizations that developers built into the testbed. This echoes some of our survey findings. It is also consistent with an observation made by a colleague that there would be great value in capturing and displaying valid human performance data in a format that (1) can be easily consumed and understood by SMEs, and (2) can be used for training, feedback, and conducting research on human-AI teaming (J. Johnston, 2023).
- It is important for the testbed to include robust data capture and storage capabilities. This was also a finding echoed in our testbed surveys.
- The testbed task should be sufficiently challenging to provide the opportunity to observe experimental effects. This is related to our survey finding that developers should be able to modulate the difficulty of the testbed task.
- It is important to ensure that the hardware and software used for the testbed has sufficient performance capabilities. This is especially true if developers will use the testbed to support machine-learning efforts.

### 3.1.2 Blocks World for Teams (BW4T)

Blocks World for Teams (BW4T) was developed by researchers from Delft University in the Netherlands and the Florida Institute for Human and Machine Cognition (Johnson et al., 2009). In it, teams of two or more humans and agents cooperate to move blocks from rooms to a drop zone in a particular order.

Figure 10 illustrates the server view of the task environment. The dark gray boxes in the top portion of the display represent rooms that players must explore to discover boxes that vary in color. The brown area near the bottom of the display is the drop area where players must bring blocks and drop them off in a particular order. The colored boxes in the lower left corner specify the order in which players must place the blocks in the drop area. The goal is to find and drop the blocks in the specified order as quickly as possible.

Individual players do not see the server view. Instead, each player has his/her/its own view of the display (see Figure 11). The black square shown between the rooms and the drop area represents the bot that that player controls. The contents of the room are displayed when the bot enters a particular room and disappear when the bot leaves the room. Players complete tasks via popup menus. The right hand portion of the client display provides "health" information, such as battery life for the controlled bot and a chat area that permits communication with other players. A given player can send messages to a specific player or all players. In the current version of the software, players can only send specific pre-defined messages.

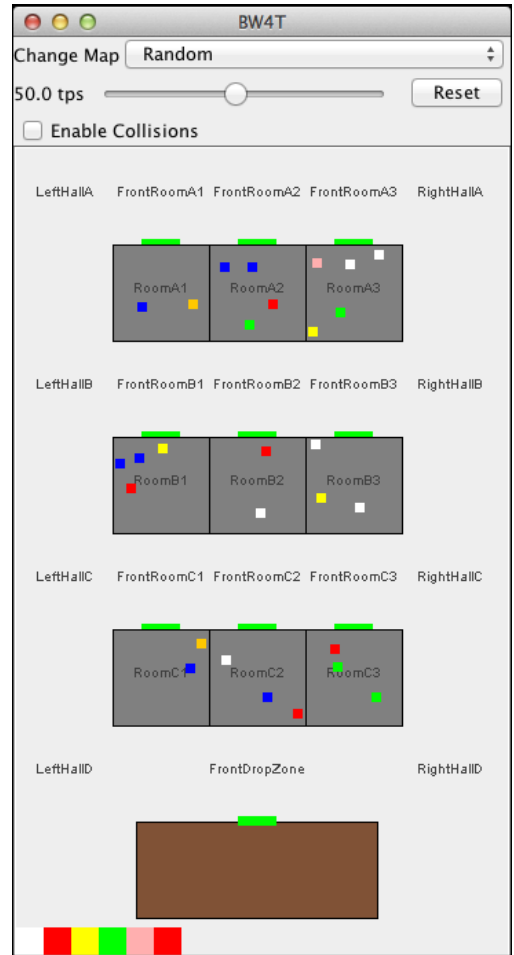

**Figure 10: Sample BW4T Server Screen (Pasman & van Riemsdijk, 2016)**

In BW4T, teams comprise at least two members. Team members can be either humans, artificially intelligent agents, or both. There does not appear to be an upper limit on team size or the number of agents that a given team can include (Pasman & van Riemsdijk, 2016; van den Oever, 2020). In the native version of the environment, cooperation is opportunistic rather than strictly required. That is, players could likely complete the task independently, but would do so more efficiently if they cooperate.

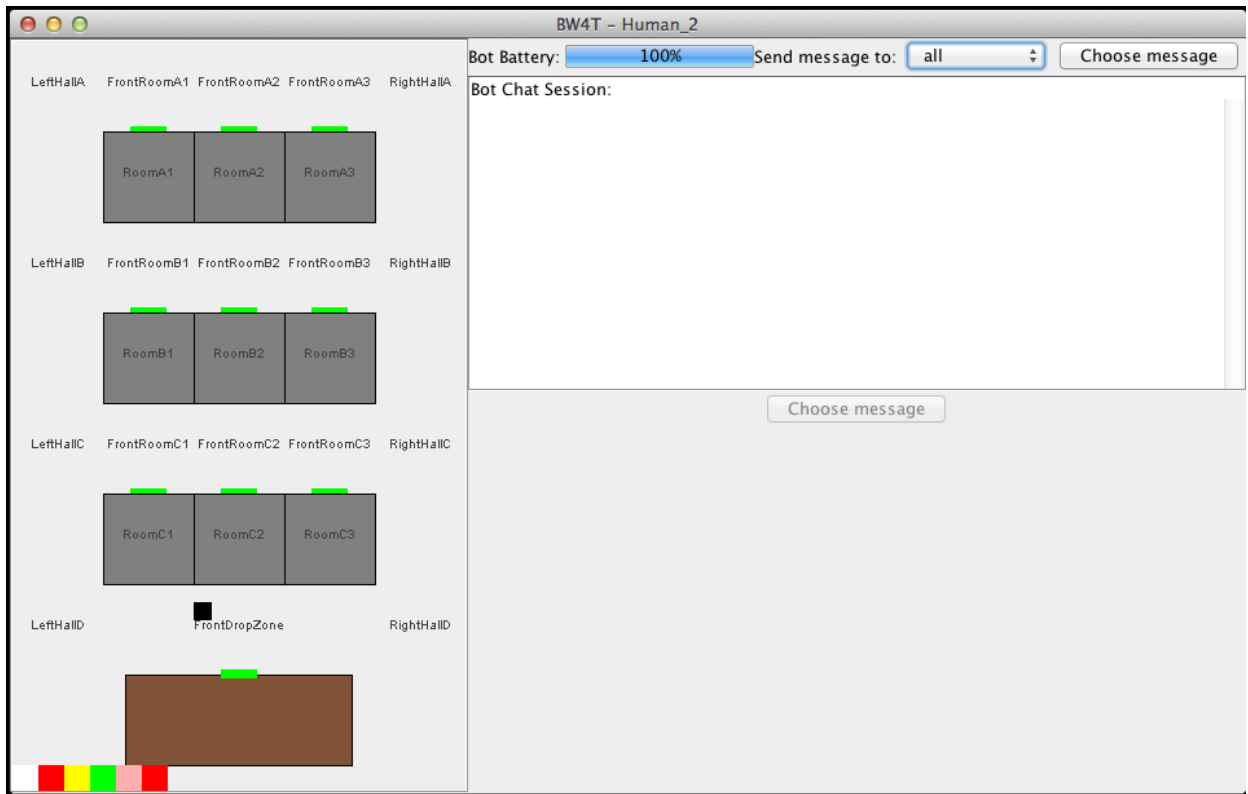

**Figure 11: BW4T Client Interface (Pasman & van Riemsdijk, 2016)**

BW4T is an open source environment (see https://github.com/eishub/BW4T) and was developed in JAVA[2]. Many researchers have modified the base version of BW4T to tailor it to specific research needs (*cf*. van Riemsdijk et al, 2012; Muise et al., 2016; and Singh, Sonenberg, & Miller, 2017). BW4T is licensed under GNU General Public License Version 3, 29 June 2007.

Researchers using BW4T develop AI agents using the GOAL programming language (Pasman & van Riemsdijk, 2016). GOAL implements the Belief-Desire-Intentions (BDI) programming paradigm (Jensen, 2021). BW4T complies with the Environment Interface Standard (EIS). EIS is a Java-based interface standard for connecting agents to controllable entities in an environment such as a game[3]. Essentially, it is an Application Programming Interface (API) that facilitates developing a connection between an agent platform and an environment.

BW4T had some capabilities that were well aligned with our previously identified desirable STE characteristics. Those capabilities were:

- The ability to collect and store basic performance data
- An open source environment with publically available code
- Support for multi-player teams that can comprise both humans and artificial agents
- A game with simple mechanics
- A game that provides an analogue for defense-related problems

---

[2] A python version of this environment may also be available. See https://matrx-software.com/. However, the maturity/stability of this code base is unknown.

[3] See https://github.com/eishub/eis

- The ability to author scenarios and modulate task difficulty

The presence of these characteristics led us to select BW4T for further analysis using the criteria outlined in Table 1. For our detailed evaluation of BW4T, see Section 4.2 and Appendix A.

### 3.1.3 MazeWorld

MazeWorld is a Unity-based testbed pioneered by researchers at Iowa State University (Cavanah et al., 2020; Francis et al., 2022) to assess the performance of teams. The researchers behind MazeWorld decided to discontinue their work before the testbed was fully developed (M. Dorneich, 2023). As a result, it is not directly applicable to our research.

However, there are some lessons learned from MazeWorld development that can improve our efforts. Specifically, Cavanah et al. (2020) and Francis et al. (2022) consciously developed MazeWorld from a measurement perspective. The research teams began by considering the teamwork and taskwork constructs that would be relevant to their research, aligned those constructs to specific behavioral indicators, and designed MazeWorld to demand those indicators. This intentional design approach is very interesting and should be considered as we move forward (see also, Ostrander, Gilbert, & Dorneich, 2019).

In addition, some of the game play mechanics (*e.g.,* the use of mazes, hidden coins, and role differentiation) may be useful to consider if we decide to tailor other testbeds (*e.g.,* BW4T) or create our own.

### 3.1.4 Reconfigurable Testbed for Teams (RT4T)

The Reconfigurable Testbed for Teams (RT4T) was inspired by BW4T (Johnson et al., 2015). The goal was to create a testbed with more generality and power than the original BW4T. The enhancements built into the JAVA-based RT4T include:

- Moving from a 2D world to a 3D world
- Moving from a third-person view to a first-person view
- Increasing relevancy (see Section 2.1) by moving from a block stacking problem to a building clearing problem
- Enhancing the code base by including data visualization utilities

The codebase for RT4T does not appear to be available to the public. As a result, it is not directly applicable to our research. However, its existence does illustrate that the BW4T codebase can form the basis for more advanced testbeds.

### 3.1.5 Others

The Multi-Agent Computing Environment (MACE) was a framework designed to support multi-agent systems research (Gasser, 1988). It was not intended to support Human-AI teaming research and is not readily available for further evaluations. As a result, we will not consider it further in this effort.

Like MACE, the Michigan Intelligent Coordination Experiment (MICE) was developed to support research in distributed AI (Durfee & Montgomery, 1989). In the case of MICE, the intent was to create a highly configurable environment that would permit the study of the coordination behavior and performance of various agent architectures. As with MACE, it was not intended to support Human-AI teaming research and is not readily available for further evaluations. As a result, we will not consider it further in this effort.

### *3.1.6 Summary*

Given the availability of the custom testbeds summarized in this section, the research team decided only to pursue an evaluation of BW4T. That evaluation can be found in Section 4.2 and Appendix A.

## 3.2 Commercial Games Used as Testbeds

An alternative to using custom-developed testbeds is to employ commercial games. While the results of our surveys indicated a strong preference for open source testbeds, our review indicated that some researchers experienced success using commercial games. This section explores that possibility.

Before considering specific testbeds built on commercial games, we want to explore some more general issues surrounding their use. Significant advantage of commercial games include the observation that they are better capitalized than other testbeds, which makes for a richer and more compelling exercise environment, and that they use the most advanced technology which enables them to offer improved performance and immersion. By using commercial games, researchers can benefit from the significant skills and investment that the developer has brought to bear on the creation of the environment. Further, commercial games often attract a large online user community with significant technical skills. The associated user community can use those skills to create extensions of the gaming environment that benefit the researcher (Leung, Diller, & Ferguson, 2005).

However, there are complications that researchers must consider when considering commercial games as the foundation of a research testbed. One is the cost of the game and any licensing restrictions that would limit their ability to use the game and report on the resultant findings. In most cases, the costs and other restrictions are not onerous, but it is important for researchers to understand their implications clearly.

We will assume that researchers will only seriously consider games that they can configure to demand varying levels of teamwork, and that the games that they consider include the modes of communication necessary to support that teamwork. An associated set of considerations is the degree to which (1) the game collects and stores sufficiently rich data elements that the research team can use to assess taskwork and teamwork, and (2) the game stores those data in a manner that the researcher can access.

Further, the research team must consider the degree to which they can modify the game to satisfy the requirements of various experimental designs, and the relative ease with which they can author novel challenges designed to assess specific research hypotheses (Leung, Diller, & Ferguson, 2005). A static gaming environment may be useful, but it is unlikely to be ideal. Many gaming environments include scripting tools that allow users to expand the game in interesting ways, but the research team must determine if they have the skills necessary to employ these tools (Carbonaro et al., 2005). Often the online community of users can be of help in this context by either providing new challenges or game features, or by answering questions that the research team may have as they work with the game (Leung, Diller, & Ferguson, 2005; Carbonaro et al., 2005). For this reason, the vitality of the game's user community is another consideration when evaluating the suitability of a commercial game as the foundation for a testbed.

The last set of considerations have to do with experimental throughput. Some commercial games are complex and require a long time to develop basic competency. Until that level of competency is achieved, any differences associated with the experimental intervention may be swamped by the noise generated by less than competent game play. In a related vein, the challenges presented by many games can take a while to complete. While this may create a rich body of teamwork and taskwork performance data, it can also limit the throughput of the research. In many cases, it is helpful to select games that are easier to master and that have challenges (*i.e.,* scenarios, chapters, levels, etc.) that are shorter in duration (Leung, Diller, & Ferguson, 2005).

We reviewed ten environments within this section:

1. Overcooked! 2
2. Overcooked!-based "Onion Soup" testbed variation (Carroll et al., 2019)
3. Overcooked!-based "Apple Juice" testbed variation (Gao et al., 2020)
4. Overcooked!-based "Salad" testbed variation (Wu et al., 2021)
5. Overcooked!-based "Gather Ingredients" testbed Variation (Song et al., 2019)
6. Overcooked!-based "Sushi" testbed variation (Buehler et al., 2021)
7. Situation Authorable Behavior Research Environment (SABRE; Leung et al., 2005)
8. Artificial Social Intelligence for Successful Teams (ASIST) Study 3 Testbed (ASIST Saturn+)
9. ASIST Study 4 Testbed (ASIST Dragon)
10. Black Horizon

### 3.2.1 Cooking with Humans and Autonomy in Overcooked! 2 for studying Performance and Teaming (CHAOPT)

*Overcooked! 2* is a commercial game that positions players in a restaurant kitchen and asks them to prepare meals in a specified time-constrained sequence. The goal is to deliver meals correctly and in a timely fashion; failing to do so lowers players' scores. There is significant time pressure built into the game, and interdependence increases as the players progress through levels (Bishop et al., 2020; Tossell et al., 2020).

The Cooking with Humans and Autonomy in *Overcooked! 2* for studying Performance and Teaming (CHAOPT) testbed employs an unmodified version of the game that will not be suitable for our research (Bishop et al., 2020). Critically, as a commercial product, the game cannot be modified and does not interface with intelligent agents (instead, the researchers used a Wizard of Oz paradigm to simulate the functioning of agents).

It is worth noting, however, that a number of researchers have developed custom testbeds strongly and explicitly influenced by the commercial version. One or more of these customized versions may be an appropriate testbed for our research.

Carroll et al. (2019) created one such variation to support their investigation of the relative effectiveness of various coordination strategies within a cooperative game. Their software is available via GitHub.[4] In their version, the human and agent must cooperate to make and plate soup. That is, they have to get onions from an onion dispenser, put three onions in a pot, wait a certain amount of time for the soup to cook, get a dish from a dish dispenser, put the soup in the dish, and move the dish to a serving location. The goal is to distribute as many orders of soup as possible within a given time period. Carroll et al. (2019) created five separate kitchen layouts to create varying levels of required coordination. It appears that this version of the software only supports two players (notionally, one human and one agent).

Gao et al. (2020) created a similar environment to support their research that focused on the value of estimating a user's mental model in order to recognize and respond to the presence of coordination disconnects. Their game, developed using the Unreal simulation engine, focused on making apple juice.

---

[4] GitHub - HumanCompatibleAI/overcooked_ai: A benchmark environment for fully cooperative human-AI performance.

The software is available on GitHub[5]. Within this game, the human and agent must gather three apples, slice them, move the slices to the juicer, run the juicer to create apple juice, and then pour and deliver the apple juice. As with Carroll et al. (2019), the Gao et al. (2020) game seems tailored to two players (one human and one agent). It seems that only one kitchen was used, and the level of difficulty, interdependence, time pressure, *etc*. was not varied.

Wu et al. (2021) created another similar task environment in support of their research on multi-agent coordination. As with the earlier examples, the goal is to complete a given recipe as quickly as possible, with a time-out defined at 100 time steps. In this case, the recipes all appear to be various simple variations of a salad (*e.g.,* the presence or absence of lettuce or tomato in a chopped or unchopped form). The Wu et al. (2021) game seems to support three or more kitchen configurations (no divider, partial divider, complete divider); multiple recipe variations; and two or three players. However, the description of the environment was not detailed, so these assumptions needed to be confirmed via code inspection. The software for this testbed is available on GitHub [6].

In their investigation of hierarchical reinforcement learning, Song et al. (2019) created another specialized version of *Overcooked*. In their version, an agent had to gather various ingredients in a specified order and deliver them to the chef. To do so, the agent had to learn to move each of its four legs independently in the desired direction (which differed for each ingredient needed in each episode). The research team shared their software on GitHub [7]. A significant disadvantage of this version of the game for our purposes is that it is limited to one player and that player is likely an agent.

Buehler, Adamy, and Weisswange (2021) were also inspired by *Overcooked*. In their investigation of communication strategies that could facilitate cooperation within human-robot teams, these researchers modeled a sushi-making game on *Overcooked*. The research team created task complexity by requiring players to prepare six different recipes. In developing the game, Buehler, Adamy, and Weisswange (2021) established the need for coordination, communication, and planning by only allowing one item at a time at all locations except the assembly area. Unfortunately, two issues limit the relevance of their game to our research. First, they did not provide a link to their code base. Second, their game involved only two players.

Our review of *Overcooked* and its non-commercial analogs suggested various issues that we may want to consider in the development or selection of a testbed. One important issue is a consideration of the level(s) and type(s) of fidelity that are required for our research. The various types of fidelity can be seen as related to the concept of "relevancy" introduced in Section 2.1. In considering simulation-based systems, attention is often focused on physical fidelity (*i.e.,* the extent to which the simulated environment is a veridical recreation of the actual environment). However, Bishop et al. (2020) and Tossell et al. (2020) remind us that cognitive fidelity (*i.e.,* the extent that the simulator reproduces the types of cognitive activities involved in the real-world task) and related concepts such as psychological or conceptual fidelity may be just as important as physical fidelity for certain types of research questions. For instance, the surveys conducted with SMEs and members of the research community revealed the desire for tasks to reflect similar information processing challenges to certain military operations of interest. A task with psychological fidelity to an operational task of interest might require complex information processing, pose situation awareness challenges, or require divided attention. Tasks that pose these challenges need not necessarily do so with physical fidelity to military operations.

---

[5] XCooking (xfgao.github.io)
[6] https://github.com/rosewang2008/gym-cooking

[7] GitHub - YuhangSong/DEHRL: Diversity−Driven Extensible Hierarchical Reinforcement Learning. AAAI 2019.

Psychological fidelity without physical fidelity is more likely to be acceptable for exploratory research; applied research focused on a specific use case might emphasize physical fidelity to a greater extent.

Another issue that may be of interest is how subjective measures of team and environment perceptions could be combined with measures defined based on various game mechanics (Bishop et al., 2020; Rosero, Dinh, de Visser, Shaw, & Phillips, 2021[8]).

Overall, there were a number of features of the *Overcooked!*-based testbeds that led us to investigate them more fully in our testbed assessment. Those features were:

- Publically-available code for some non-commercial testbed variations
- Ability to define various aspects of scenarios
- Designs that are conducive to the integration of synthetic agents

### 3.2.2 Situation Authorable Behavior Research Environment (SABRE)

Leung, Diller, & Ferguson (2005) created the Situational Authorable Behavior Research Environment (SABRE) to explore the viability of using commercial game technology to study team behavior. Their goal was to capitalize on the immersive and motivating nature of games to create an environment in which participants would perform more naturally.

Leung and her colleagues based SABRE on *Neverwinter Nights*™, produced by Bioware. *Neverwinter Nights* is a role-playing game[9] based on Dungeons and Dragons. While the source code for *Neverwinter Nights* is not available, it does include an API, editing tools, and a scripting language that allow users to extend the base game. Further, an active user community uses the tools to create variations and extensions that are generally available for use. For example, in developing SABRE, Leung, Diller, & Ferguson (2005) used a community-developed cityscape as opposed to the default "medieval" setting.

Leung, Diller, & Ferguson (2005) describe gameplay in this way:

"In *Neverwinter Nights* players can move around the world, pick up and use items, go into buildings, read maps and signs, and interact with other characters. They can communicate with other players through free form typed text, or engage synthetic characters through dialog menus. Players can use a journal and map, as well as a variety of items within the environment. They may also attack targets, but do not have much detailed control over fighting actions."

As a concrete example of the use of SABRE in research, the North Atlantic Treaty Organization (NATO) Research and Technology Organization (RTO) Human Factors and Medicine Panel partnered with the NATO Allied Command Transformation (ACT) Futures and Engagement Concept Development and Experimentation to research optimized adaptations to cultural differences that affect teaming. This was done through a project entitled Leader and Team Adaptability in Multinational Coalitions (LTAMC). The LTAMC research included a number of studies on multi-cultural teaming using the SABRE/*Neverwinter Nights* testbed (NATO RTO, 2012). Researchers assembled teams of four, with one teammate randomly selected for the leader role. The team's goal was to search a virtual city (an urban overlay developed for the *Neverwinter Nights* game) and locate hidden supplies of weapons while earning (or losing) good will with the non-player characters (NPCs). Good will was tracked through points that were accumulated by finding the weapons caches, establishing good relationships with NPCs, and performing side-quests (Warren, 2012). The team had the opportunity to assign individual responsibilities and create communication and search plans. All communications occurred via a chat box, and team members could broadcast messages widely to teammates in the immediate vicinity or send them long distance to individual team members. The testbed collected data on weapons cache finds, chat counts, information sharing, map tool markings, good will points, and situation awareness.

---

[8] This study also noted differences in communication patterns within all human and Human-AI teams. This suggests that it may be beneficial to include communication best practices within a notional team-training curriculum to see if these differences can be minimized and whether doing so improves the performance of Human-AI teams.

[9] In role-playing games, each player assumes the role of a character who can interact within the game's imaginary world. These games often present challenges that one or more player characters must overcome to advance.

A notably positive feature of using this testbed for this research effort was the game's unstructured environment (Warren & Sutton, 2012). According to the researchers conducting studies as a part of the LTAMC effort, *Neverwinter Nights* offered a creative environment with many opportunities for "self-determinism." In other words, the game offered a rich decision-making environment for teams. In addition to the freedom permitted within the *Neverwinter Nights* world, the game was also highly immersive, and participants were more involved in the game than with the nature of the study being conducted. Finally, "dungeon-master mode," allowed experimenters to access the game environment via an invisible avatar and intervene in the game space if needed.

While the SABRE testbed is not available to us, Sonalysts has the technical ability to develop a similar extension. Initially establishing a testbed using *Neverwinter Nights* would likely involve a significant one-time effort, with less intensive ongoing work to maintain the testbed (although perhaps more work than using an open source environment). The features of *Neverwinter Nights* that make it an intriguing option to investigate further include:

- Available API for enhancing game functionality
- The ability to log and assess many types of player actions
- Regular software updates[10]
- Support for teams with multiple teammates and various defined roles (and scripting tools to facilitate flexibility in team size and structure)
- Flexibility to easily create and modify task features/challenges (such as scenario difficulty, information processing complexity, etc.) with game editing tools
- A rich user community

### 3.2.3 *The Artificial Social Intelligence for Successful Teams (ASIST) Minecraft Testbeds*

In the same way that the CHAOPT and SABRE teams built their testbeds around commercial games (*Overcooked! 2* and *Neverwinter Nights*, respectively), the Artificial Social Intelligence for Successful Teams (ASIST) program (Elliot, 2019) aimed to study Artificial Social Intelligence (ASI) as an advisor in Human-AI teaming. The performers started from the open sourced research platform Microsoft Malmo[11] and built several versions of *Minecraft* testbeds (Kickoff: ASIST Sparky; Study 1: ASIST Falcon; Study 2: ASIST Saturn; Study 3: ASIST Saturn+; Study 4: ASIST Dragon). Of these, ASIST Saturn+ is best documented so far. More information about all testbeds will be available the ASIST website[12] in the future.

ASIST Minecraft testbeds aim to serve multiple research purposes.

- Support the development of ASI agents to improve teamwork and its effects through impactful interventions.

---

[10] The development team would benefit from software updates that have occurred throughout the years (see https://store.steampowered.com/app/704450/Neverwinter_Nights_Enhanced_Edition/).
[11] https://blogs.microsoft.com/ai/project-malmo-lets-researchers-use-minecraft-ai-research-makes-public-debut/
[12] artificialsocialintelligence.org

- Support the development of real-time Analytical Components (ACs) that provide social science-based measures to drive ASI or assess its impact.
- Support the training and evaluation of ASI advisors in comparison to ground truth and each other.
- Accelerate data collection for training and evaluation by streamlining experiment administration, shortening experimental sessions, and collecting large-scale human subject data for advanced machine learning modeling.
- Test the generalizability of findings across different experimental settings.

The ASIST Study 3 testbed (hereafter, ASIST Saturn+) presented an urban search and rescue (USAR) task scenario (Huang et al., 2022a). Using this scenario, researchers collected multimodal data (*e.g.,* videos, audios, testbed messages, and surveys) from 113 teams. Each team comprised three human members and zero or one ASI advisors. The data are available for public access (Huang et al., 2022b). The ASIST Saturn+ testbed software (other than the commercial version of *Minecraft*) is open source and available on Gitlab[13]. Researchers with strong programming skills can spin up this testbed independently, but further support may need to wait until after the launch of the ASIST Study 4 scalable testbed (hereafter, ASIST Dragon) in fall 2023. The ASIST Dragon testbed code will be open source on Gitlab at that time. Researchers who wish to use ASIST Dragon will have extensive technical support from 2024 to 2025.

The ASIST Dragon testbed extended the infrastructure of the ASIST Saturn+ to support admin-less parallel collection of data from multiple teams. It will use a scaling architecture that dynamically increases the number of servers based on the number of players in an online waiting room. Participants in the study connect to the centralized waiting room, which handles registration, consent, and pre-trial surveys. Teams are formed automatically and then connected to individual game servers, which handle initialization of virtualization containers for each agent. For Study 4, ASIST Dragon will be configured to use a "First-In, First-Out" team formation approach, but this will be configurable to other methods for team formation. Data from each trial is recorded to a centralized database so that agents can access aggregated data across plays. When ASIST Dragon launches in fall 2023, it is expected to have the following features:

1. Three-person teams with minimal domain-specific expertise
2. Optional interdependent team tasks (e.g., tool purchase options create additional interdependence opportunities in bomb disposal, fire extinguishing, and rescuing teammates)
3. A recon phase to warm up and get familiar with the environment
4. Cognitive team tasks (e.g., planning and negotiation on resource distribution and strategy)
5. Natural language chat text and machine readable cognitive artifacts
6. Real-time ASI advisor interventions via Minecraft chat based on real-time analytical component
7. Real-time feedback to ASI advisors' interventions via pre-determined response options
8. Fire and bomb explosion related injuries and risks
9. Perturbations (e.g., unexpected fire) to disrupt normal tasks
10. Changeable task environment (e.g., bomb layouts, regions simulating desert, village, or forest)
11. Intake surveys and post-trial surveys
12. Multiple servers available 24/7 without administrative effort for data collection
13. Supports participants engaging in repeated trials
14. A primary data output of testbed messages, including the timestamps, events, ASI interventions, and surveys, etc.

---

[13] https://gitlab.com/artificialsocialintelligence/study3

15. Social science implemented in Analytic Components drives ASI Machine Theory of Mind (MToT), which in turn drives ASI interventions, influences team process, and potentially influences mission effects

Both ASIST Saturn+ and ASIST Dragon testbeds present solid teamwork task scenarios designed by leading experts in teamwork research based on numerous discussions, prototypes, and pilot tests. Researchers can modify all the scenarios to address additional research questions. Below is a brief description of ASIST Saturn+ task scenarios. Detailed descriptions can be found in Huang et al. (2022a).

In ASIST Saturn+, a team of three players is responsible for searching for, stabilizing, and transporting victims to appropriate safe zones in a collapsed building in two 17-minute missions. The victims suffered injuries at various levels of severity, and require different levels of treatment (see Table 3). Players assume one of three roles within the team: Medic, Engineer, or Transporter. Each role possesses unique knowledge and capabilities, while also being linked with shared skills (*e.g.,* the ability to transport stabilized victims at various speeds). The Medic's primary role is to stabilize victims, with the special knowledge of victim injury types. The Engineer's primary role is to clear rubble from pathways and around victims and to rescue teammates if they are trapped by fallen rubble in "threat" rooms, with the special knowledge of where the threat rooms are. The Transporter's primary role is to move stabilized victims to the appropriate safe zones at the fastest speed, with the special skill of detecting victims in rooms at the door without entering the room. The team's goal is to maximize its point total by rescuing an optimal collection of victims.

Table 3: ASIST Saturn+ Victim Profiles

| Victim Type | Attributes | Implications |
|---|---|---|
| A | Minor wounds, such as abrasions | - Takes 3 seconds to stabilize<br>- Worth 10 points, each<br>- Must be transported to Safe Zone A |
| B | Moderate wounds, such as fractures | - Takes 3 seconds to stabilize<br>- Worth 10 points, each<br>- Must be transported to Safe Zone B |
| C | Critical, life-threatening wounds | - Must be awakened prior to stabilization<br>    ○ Task requires a Medic and one other player<br>- Can be stabilized after awaking<br>    ○ Takes 3 seconds to stabilize<br>- Worth 50 points<br>- Must be transported to Safe Zone C |

As noted above, ASIST Saturn+ defines three roles with unique responsibilities and capabilities. That number seems fixed, but could potentially be enlarged by including several Search and Rescue teams in a "team of teams" format. In the current implementation, the three human players conduct the tasks, with an advisor (*i.e.,* a human, or one of six ASI agents) giving advice to improve the human teamwork. This structure provides an encouraging indication that it may be possible for researchers to assign AI agents to specific roles.

The ASIST Saturn+ architecture includes ASI agents as additional "services" and integrates them by establishing a connection to the message bus. This allows the agents to subscribe to messages produced by other testbed components. To facilitate this process, the ASIST program performers have produced a software library (ASISTAgentHelper) that automates many agent infrastructure tasks. In addition, two sample agents were provided for developers to use as design models. The final testbed included six ASI agents.

The ASIST Saturn+ scenarios include several features that developers could use to control scenario difficulty. Two perturbations can be introduced to increase the complexity of a scenario. One is the occurrence of a structure collapse that introduces new rubble that the engineer must clear. The other is the signal loss of GPS-based maps. Additional examples that developers could adjust include:

- The complexity of the tasks (*e.g.*, additional tasks)
- The capabilities of the roles: infinite tools or tools that can be replenished
- The complexity of the task environment (*e.g.*, the building layout and environmental features)
- The number and types of victims
- The victims' locations
- The number and location of blockages and threat rooms
- The presence or absence of perturbations

ASIST Saturn+ includes robust data collection and performance measurement features. All components within the testbed publish data on the primary MQTT message bus, and Analytic Components (ACs) use the published data to provide specific measurements and register results on the message bus. Table 4 lists some of the measures that are "built in" to the testbed. Moreover, developers can create novel ACs as needed.

**Table 4: Key ASIST Saturn+ Performance Metrics**

| Measure | Description |
| --- | --- |
| Mission score | Total mission score for each trial |
| Subtask latency | The latency between victim discovery and victim rescue |
| Victims discovered | A report that a player has discovered a victim. An AC can sum this value over time |
| Number of victims rescued | A report that one or more players have rescued a victim. An AC can sum this value over time and can calculate a weighted sum to define the team's current mission score |
| Team member trapped/released | A report that a team member has been caught in (or released from) a trap room. An AC can calculate total time spent confined to a trap room using these reports |
| Threat room marked | The presence/absence of a threat room marker at a room entrance |
| Correct victim marks | The presence/absence of a victim marker near a corresponding victim type |

In addition to these atomic measures, the ACs can create a variety of more complex measures, including calculating various statistics. Examples include: the percentage of subtasks of teamwork tasks that are completed per trial, score per minute, communication entropy, belief entropy, total effort, percentage of time exerting role-congruent effort, percentage of SAR workload executed since mission start, goal alignment ratio, and leadership score.

The ASIST Dragon testbed embodies a bomb disposal task scenario for three-person teams supported by one of two possible ASI advisors. Dragon is designed to collect large-scale human subject team data online, with multiple servers available 24/7 without administrative effort for data collection. It will allow for studying team formation, resource allocation, risk avoidance, and teamwork behaviors. Optional interdependent team tasks (*e.g.,* tool purchase options create interdependence opportunities in reconnaissance and planning, bomb disposal, stopping environmental damage, and rescuing teammates from injuries). Detailed task scenario will be released in fall 2023 on the ASIST website.

ASIST *Minecraft* testbeds use *Minecraft* version 1.11.2, because Microsoft (the owners of Minecraft) used that version for Project Malmo, which was open source for public research. The wrapper functionality that ASIST *Minecraft* testbeds add to that foundation is open source on GitLab. The ASIST *Minecraft* testbeds use a well-structured modular architecture. The testbed site on Gitlab includes libraries and sample applications designed to facilitate development. A software engineer should review the code base to assess the complexity of extending it. The testbeds have the ability to integrate other *Minecraft* AI agents (*e.g.,* MineRL and Voyager) to create virtual agents in the current task scenarios. DARPA is encouraging Human-AI teaming researchers to use the ASIST Minecraft testbeds to explore various research questions, with limited technical support for ASIST Saturn+ and extensive technical support for ASIST Dragon after it launches.

### 3.2.4 Sonalysts-Developed Testbeds

Another related set of testbeds comprises environments that Sonalysts has or could develop internally.

One group of testbeds includes novel environments developed using Sonalysts' second-generation Simulation Engine (SEII). Although originally developed as the basis for commercial games, SEII has evolved as a modeling and simulation tool to support combat system training, system and tactics operational analysis, as well as Human-in-the-Loop (HIL) experimentation.

SEII employs a multi-threaded architecture of discrete and scalable self-contained functional modules. SEII also includes the Sonalysts-developed Global Tactical Simulation (GTS) "domain package" including a virtual global environment and library of platforms (*e.g.*, military and civilian vehicles and buildings), weapons, sensors, networks, and swappable mission payload packages. SEII/GTS facilitates the rapid development of multi-domain scenarios at the strategic, theater, operational/unit, and individual levels. Table 5 summarizes some of the advantages and disadvantages associated with using SEII to create a research testbed. For the most part, these are the advantages and disadvantages associated with using commercial game technology, in general. These tools can be used to create more "authentic" research environments, but are associated with greater costs, longer development timelines, and increased usage restrictions.

**Table 5: Advantages and Disadvantaged of using SEII to Create a Testbed Environment**

| Advantages | Disadvantages |
|---|---|
| • Ability to create simulation environments with greater relevancy (see Section 2.1) <br> • Ability to create custom-solutions tailored to specific research goals (maximizes flexibility) <br> • Likely faster/cheaper than using other commercial games | • To achieve significant relevancy (see Section 2.1), testbed development may involve significant timelines and budgets <br> • Not an open source solution, so redistribution is limited <br> • Likely requires more training and longer scenarios, limiting experimental throughput |

We have considered a variety of possible synthetic environments for our research. For example, we could base the environment in a Combat Information Center (CIC). This approach has a number of advantages for our research. In addition to the enhanced relevancy noted above, this setting would allow us to work toward a testbed that supports "teams of teams." For example, we could begin with the Air Warfare team (Anti-Air Warfare Coordinator, Identification Supervisor, RADAR Systems Controller, Missile System Supervisor, Remote Control Station, Air Intercept Controller). If needed, we could expand the testbed to include other domains, such as Surface Warfare (Surface Warfare Controller, Surface Watch Supervisor, Gunnery Liaison Officer, Gun Fire Control Systems Supervisor, Database Manager, Harpoon Engagement Planner, Engagement Control Officer) and leadership levels (Tactical Action Officer, CIC Watch Officer, CIC Watch Supervisor, Combat Systems Coordinator, Combat Systems Officer of the Watch, Communications Watch Officer, Tactical Information Coordinator).

Another option would be to base the testbed in an organization such as an Air Operations Center (AOC). This approach would allow the teams-of-teams approach, and has the added advantage of being a reasonable surrogate for a JADC2 operating structure. Other options that we considered included (1) building an incident-response game that mirrored the team organization within an incident response center or (2) approximating a JADC2 framework by developing a testbed environment based on the Combatant Command staff structure (*i.e.,* the various J-level directorates).

An alternative to SEII for testbed creation was to use the foundation provided by Sonalysts' Standard Space Trainer (SST). SST is a simulation-based practice environment that presents each learner with controls/displays that replicates his/her operational console to the fidelity required to satisfy training objectives. Scenarios are developed and delivered to present learners with events and other trigger conditions associated with operational tasks. An instructor monitors one or more learners as they complete these tasks. To help with this process, SST provides the ability to facilitate reliable performance assessment and to "drop in" on learners virtually to observe their performance. Instructors can communicate with learners to provide real-time assessment and coaching, and together they can review practice sessions within a replay environment to reinforce key teaching points. SST supports individual, team, and team-of-teams training.

SST consists of two major components: the runtime environment and the Software Development Kit (SDK). The SST runtime environment provides the common training features that are constant across the individual training systems. One way of thinking about this is that the runtime environment provides the DNA that gives all SST trainers a strong family resemblance. These common components include the supports for instructor-supported practice, assessment, and feedback; re-usable and expandable models; and data collection and analysis tools. The SDK and API allow vendors to build custom applications that use and expand on the foundation provided by the runtime environment. Vendors use the SDK to develop a training system for a particular operational system. Using these tools, developers replicate and instrument the User Interfaces that operators use to complete their tasks and develop any special-purpose models. The SDK promotes adherence to SST architectural and training standards. One of the important advantages of using the SDK to produce the training systems is that it allows multiple training systems to be deployed on the same hardware.

Given its authorability and rich data capture abilities, the SST framework provides an attractive testbed option. However, it suffers from many of the same limitations of a testbed developed under SEII. Specifically, we could use it to develop more immersive scenarios, but at the cost of longer development timelines and increased budgets. The more immersive environments may take longer to master, thus limiting throughput. In addition, SST in not open source and cannot be freely distributed.

Another Sonalysts-developed game that could serve as the foundation of a testbed is "Black Horizon," a spacecraft engagement simulation with a storyline, quick mission cycling, realistic graphics and physics, and instructional scaffolding that trains learners to manage the challenges of maneuvering a space vehicle in a contested environment. Missions in Black Horizon are designed specifically to exercise skills learners are working on in training. Mission goals are tied to particular learning objectives, and accomplishment of those goals are indicative of mastering learning objectives.

In Black Horizon, each learner plays as a member of a (fictional) secret space peacekeeping organization. Players learn to control an individual satellite and to coordinate formations of satellites with different sensor, communication, and weapon capabilities. The learner in the game is trained to perform increasingly complex orbital maneuvers as they advance through different levels. Two multiplayer environments are available in Black Horizon; one allows multiple players to perform the same mission on a single IP address, and the other is online, allowing players to locate and engage other players on the internet. Some of the characteristics of Black Horizon that make it an attractive potential candidate for the testbed include:

- Flexibility of instrumentation for data collection
- Tasks with high-fidelity to defense-related operations

### 3.2.5 Summary

The research team decided to pursue an evaluation of the various *Overcooked!*-based testbeds developed by researchers, as well as *Neverwinter Nights*, ASIST Saturn+, and Black Horizon. Those evaluations can be found in Section 4.2 and Appendix A.

### 3.3 Open source Games Repurposed as Testbed

Open source games may represent a compromise between custom-built researcher testbeds (see Section 3.1) and commercial games (see Section 3.2). These games are generally developed by enthusiasts as a hobby and are maintained by the community. They often take longer to mature than commercial games, and can lack the production values of high-end games (but generally exceed those of custom-built testbeds). They are published under an open source license such as the MIT License, GNU General Public License, or the BSD License. The combination of a sophisticated gaming experience and extensible code base makes them an attractive option for our Human-AI teaming testbed.

As an example of such a testbed, consider *Spring* (Bakkes, Spronck, & van Den Herik, 2008; 2009). Spring is an open source real-time strategy game (see https://springrts.com/ and https://github.com/spring/spring). Like most games in this genre, players compete against each other by gathering resources, building structures, and developing military units to protect and capture land and resources. In *Spring*, the game ends when one player/team destroys the opponent's "Commander" unit.

Other testbeds that we may consider in this category include:

- Globulation 2
- Unvanquished
- Netrek
- Crossfire

A challenge with using these testbeds is that it is not clear whether they would meet industrial security standards, such as those imposed by Cybersecurity Maturity Model Certification (CMMC). With open source software, any developer can add to or edit the code, which potentially presents security risks. Research into the risks associated with open source software is ongoing. Therefore, the research team did not select any of these testbeds for evaluation at this time.

## 3.4 Summary

There are myriad benefits and drawbacks to any of the software options outlined in this section. Open source testbeds provide a free, publically available task environment that could be modified to develop a teaming testbed. However, they could present significant security risks and may not be suitable for our use. For simple open source testbeds, one possible approach to resolving security risk is to create our own version of a testbed based on the open source game (as a number of researchers did with their variations on the commercial game *Overcooked!*). Commercial games require a baseline expense to use, which prevents us from creating a truly open source testbed environment (an important feature for the research stakeholders that we surveyed at the outset of the effort). However, they also tend to be more engaging than a game we could develop as a stand-alone product, and available APIs and authoring tools make them amenable to fundamental teaming research.

Testbed selection for formal evaluation was based on available information about the testbed and a preliminary review of implementation factors (testbeds that presented implementation "nonstarters" were not considered). The research team selected the following testbeds for further evaluation:

- Blocks World for Teams (Johnson et al., 2009)
- Black Horizon
- Artificial Social Intelligence for Successful Teams (ASIST) Study 3 testbed (ASIST Saturn+; Huang et al., 2022a)
- ASIST Study 4 testbed (ASIST Dragon)
- *Overcooked! 2*
- *Overcooked!*-based "Gather Ingredients" testbed variation (Song et al., 2019)
- *Overcooked!*-based "Apple Juice" testbed variation (Gao et al., 2020)
- *Overcooked!*-based "Onion Soup" testbed variation (Carroll et al., 2019)
- *Overcooked!*-based "Sushi" testbed variation (Buehler et al., 2021)
- *Overcooked!*-based "Salad" testbed variation (Wu et al., 2021)
- *Neverwinter Nights* (SABRE testbed; Leung et al., 2005)

After identifying these candidates, the research team subjected them to a formal structured evaluation. The evaluation process and its results are discussed in the next section. The full evaluation of these testbeds can be found in Appendix A.

## 4 QUANTITATIVE ASSESSMENT OF TESTBED ALTERNATIVES

### 4.1 Method

In order to conduct a quantitative assessment of testbed alternatives, we must define explicit testbed criteria and use those criteria to create an evaluation tool. We began this process by reviewing the desirable testbed features derived from the research stakeholder survey data (found in Table 1 of Section 2.2). The next step was to establish the relative value of each feature. If all features are weighted equally on the evaluation tool, then the value of each testbed is the sum of quality scores. However, some qualities are more important than others when selecting a testbed to meet research needs for Human-AI teaming. To reflect these differing priorities, we developed weights for the criteria to be incorporated into the testbed evaluation tool. We established some of these weights based on project requirements (such as cost and speed of implementation). We based other weights on survey data to reflect stakeholder input. That process is described below.

We designed and distributed surveys for two groups of stakeholders: researchers who conduct studies on Human-AI teams, and military SMEs with experience in JADC2 or other command and control type operational settings. These surveys had similar organizations, in that they combined open-ended questions about ideal STE features and Likert-style feature-rating questions. We used responses from both of these question types to weight testbed criteria.

The research team started with the 23 qualities that we introduced in Section 2.2. To prioritize these features, we used three factors to contribute to weights:

1. Likert item strength
2. Overlap with researcher and SME survey themes
3. Internal research priorities

In both the researcher and SME stakeholder surveys, respondents ranked their agreement with the inclusion of a number of features in the STE. We wanted to assess the strength of agreement with those features as a factor for weighting criteria. We mapped each of the relevant Likert items from both surveys to the 23 qualities defined in Section 2.2, calculated a mean agreement score for each of these qualities, and converted that score to a 1-3 scale. Note that not all qualities had Likert items that mapped to them, because some of the qualities were added based on research project priorities rather than strictly survey data. Therefore, some of the qualities scored "0" on this scale. The Likert strength score for each quality can be found in Table 6.

Both researcher and SME surveys also had a number of open-response style questions that prompted respondents to list ideal STE features. For both sets of open-ended responses, we conducted an initial thematic analysis of responses to create a master list of recurring ideas across answers. To apply these themes to the evaluation effort, the themes from these lists were mapped (where applicable) to the 23 STE qualities identified in Section 2.2 (note that many of these qualities were derived from the researcher themes, so more research themes mapped to qualities than SME themes). We created a 1-3 scale to reflect the amount of overlap of each STE quality with researcher and SME thematic analyses, where "1" represented no overlap with either; "2" represented overlap with responses from one of the stakeholder groups; and "3" represented overlap with both stakeholder groups. The scores on this overlap scale for each feature are shown in Table 6.

Table 6: Testbed Evaluation Criteria, Criteria Weights, and Features

| Feature | Likert Overlap Score | Themes Overlap Score | Internal Priority Ranking | Evaluation Criteria | Evaluation Criteria Z-Score | Final Weight |
|---|---|---|---|---|---|---|
| Can be instrumented to collect a variety of data, including individual and team status. | 2.11 | 3 | 6 | Data Collection/ Performance Measurement | 1.71 | 3 |
| Collect process measures. | 2.54 | 3 | 9 | | | |
| Collect outcome measures. | 0 | 3 | 17 | | | |
| Implementation Cost | 0 | 2 | 1 | | -0.808 | 3 |

| Feature | Likert Overlap Score | Themes Overlap Score | Internal Priority Ranking | Evaluation Criteria | Evaluation Criteria Z-Score | Final Weight |
|---|---|---|---|---|---|---|
| Speed of Implementation | 0 | 1 | 2 | Implementation Factors | | |
| Support teams with multiple defined roles (4-12). | 2.24 | 3 | 7 | Teaming Factors | 0.742 | 2 |
| Allow each role to be filled by a human or agent. | 1.76 | 3 | 8 | | | |
| Distinguish between the source and intended recipient for each communication act. | 0 | 2 | 23 | | | |
| Ease the creation of unique scenarios aligned with research objectives. | 1.93 | 3 | 11 | Scenario Authoring | 0.406 | 2 |
| Facilitate the definition of linkages among competencies, scenario events, and assessment routines. | 1.95 | 2 | 19 | | | |
| The task should include adjustable levels of information density, signal to noise ratio, and data reliability. | 0 | 1 | 3 | Task Features | 0.333 | 2 |
| Tasks should impose conflicts among individual goals and team goals. | 0 | 1 | 4 | | | |
| Task demands should replicate or impose information processing demands analogous to real-world tasks. | 1.92 | 3 | 5 | | | |
| Tasks should require communication among teammates (*e.g.*, spoken, text, chat). | 1.90 | 3 | 10 | | | |
| Task should be easy to learn but potentially challenging. | 1.45 | 3 | 12 | | | |
| Tasks should have adjustable levels of difficulty, uncertainty, time pressure, *etc*. | 2.15 | 3 | 13 | | | |
| Tasks should require significant coordination/interdependency. | 2.15 | 2 | 22 | | | |

| Feature | Likert Overlap Score | Themes Overlap Score | Internal Priority Ranking | Evaluation Criteria | Evaluation Criteria Z-Score | Final Weight |
|---|---|---|---|---|---|---|
| Include "quick look" visualizations of individual variables and variable pairs. | 1.22 | 2 | 14 | Data Processing | -0.37 | 1 |
| Facilitate data export. | 2.33 | 2 | 18 | | | |
| Modular/Flexible | 0 | 2 | 16 | System Architecture | -0.53 | 1 |
| Open source | 2.13 | 2 | 20 | | | |
| Permits collaboration across labs. | 2.19 | 2 | 21 | | | |
| Facilitate integration of agents with differing capabilities. | 1.86 | 2 | 15 | Agents | -1.48 | 1 |

We used our internal research objectives to prioritize qualities further. First, we ranked the 23 qualities in relative importance, given our internal priorities. If an item appeared in the top third of the list, we *added* 1/3 of a point to its score. If an item appeared in the bottom third of the list, we *subtracted* 1/3 of a point from its score. No changes were made to the nine qualities in the middle. From there, an "overall weighting score" was calculated for each quality by adding the Likert strength score with the overlap score, dividing by 6, then adding the point value associated with the internal research objectives. These scores were then used to calculate an "overall weighting score" for each of the broad criteria categories outlined in Table 1. Z-scores were calculated for these criteria groupings to look at the distance of each group from the mean. Anything with a Z-score above one was given a weight of three. Categories with positive Z-sores that were below one were assigned a weight of two, and categories with negative Z-scores were given a weight of one. These final weight scores can be found in Table 6.

The process described above worked well to produce weights for seven of the eight evaluation criteria. The only criterion with a score that was inconsistent with our needs for this project was "Implementation Factors." Because we did not explicitly create Likert items about implementation features, and our respondents did not spontaneously list them in their responses to open-ended questions, this criterion received a very low score. However, cost and schedule are always important, particularly for an internally funded effort. Therefore, we "overrode" the scoring formula to reflect the practical significance of implementation factors, and changed the weight of that criterion from a "1" to a "3".

Final criteria weights in Table 6 provide values for weighting factors for the testbed evaluation tool (in Table 8). The testbed is rated on a scale from 1-10 based on the extent to which it meets the criteria (descriptions of criteria were developed using descriptions of associated features listed in Table 1). This is assessed using a combination of (1) the proportion of features within each criteria met by the testbed and (2) the general feeling of the rater that the criteria is met. For example, the "System Architecture" criteria has of three features: modularity/flexibility, an open source framework, and permitting collaboration across labs. If a testbed has two of those three features, one might give the testbed a 6 or 7 to reflect the proportion of features met. Whether the testbed receives a rating of 6 or 7 may depend on how well the rater feels the other features are met by the testbed. Since there is an element of subjectivity in this process, the research team used two independent raters to conduct assessments of testbeds, reviewed those scores, and worked to resolve any noticeable discrepancies among the scores.

To help guide evaluation of each of the established criteria, the researchers developed guidelines for evaluation presented in Table 7.

Table 7: Scoring Standards for STE Evaluation Criteria

| EVALUATION CRITERIA | DESCRIPTION | SCORING STANDARD | | |
|---|---|---|---|---|
| | | 0-3 | 4-6 | 7-10 |
| DATA COLLECTION / PERFORMANCE MEASURES | The testbed is instrumented to collect data on individuals and teams for both outcomes and processes. | The testbed collects a few fixed data points. | The testbed collects a variety of outcome and process measures. | The testbed includes rich instrumentation and likely supports the ability to define unique measures of interest. |
| IMPLEMENTATION FACTORS | Cost of implementation is low and speed of implementation is fast. The testbed is directed toward data collection and analysis rather than specific tool development. | The software architecture and/or problem space is complex. This will lead to long/costly development and/or modification timelines. | The architecture is reasonable, but various factors suggest that a good deal of software engineering and/or subject-matter expertise will be needed to develop/modify specific scenarios. | The system has a well-considered modular architecture. The problem space and/or authoring tools are simple enough to enable speedy and cost-effective development/modification. |
| TEAMING FACTORS | The testbed supports teams of 4-12 members, flexible assignments of roles to humans or agents, and measurement of communication between teammates. | The size of the team is small and fixed. Roles are not differentiated. Autonomous agents are not incorporated. Communication between teammates is not recorded. | Team size is appropriate (*e.g.,* 4-6) and autonomous and human teammates can be assigned to multiple roles. Team communication is recorded. | Team size and configuration (autonomous/human teammates) is very flexible and roles are (or can be) clearly defined. All modes of communication are recorded for evaluation. |

| EVALUATION CRITERIA | DESCRIPTION | SCORING STANDARD | | |
|---|---|---|---|---|
| | | 0-3 | 4-6 | 7-10 |
| SCENARIO AUTHORING | The STE should support the ability to create scenarios that meet our research objectives and make connections between competencies, scenario events, and measures of performance. | Scenarios are fixed or only slightly adjustable. | There is a moderate level of Authorability in the scenarios and the process seems reasonably straightforward. | The testbed includes scenario authoring tools that allow for a great deal of expressiveness and tailoring. |
| TASK FEATURES | The testbed should provide tasks with a high degree of psychological fidelity and the ability to modulate a number of task factors (*e.g.*, difficulty, uncertainty, time pressure, information density, signal-to-noise ratio, and data reliability). | The testbed tasks are fixed, artificial, and do not require a useful level of information processing. | The tasks require at least a little information processing, and some level of customization is possible to modulate basic difficulty factors. | The testbed includes tasks that demand a great deal of information processing. The tasks include a large number of features that make it reasonable to modulate important complexity dimensions. |
| DATA PROCESSING | The testbed allows for data export and data visualization. | Data export is cumbersome and/or not possible. Data visualization is not possible. | Data can be exported in conventional formats (*e.g.,* Excel or CSV). Analysis and visualization tools are not available. | There is significant support for data export in a variety of formats. Statistical analysis tools are provided. It is possible to generate data visualizations. |

| EVALUATION CRITERIA | DESCRIPTION | SCORING STANDARD | | |
|---|---|---|---|---|
| | | 0-3 | 4-6 | 7-10 |
| SYSTEM ARCHITECTURE | Testbed needs to facilitate flexible changes to the testbed as research progresses, should be open source, and should allow for collaboration among distributed teams/labs. | The architecture is fixed and monolithic. | The system architecture is reasonable and should not present insurmountable impediments to system modification. | The testbed uses a well-structured modular architecture. There are SDKs and/or other tools to support system maintenance/modification. |
| AGENTS | The STE should make it relatively easy to "plug-in" agents with a variety of capabilities. | Agents cannot be incorporated into the system. The only option is "Wizard of Oz" emulation of agents. | Agents can be developed and/or interfaced with the testbed. The interface may require significant software engineering skill. | Well-defined APIs exist to facilitate the integration of agents. |

As an example, consider the Hypothetical "Solution 1" and "Solution 2" shown in Table 8. Solution 1 meets all of the scoring criteria, and has a perfect score, while "Solution 2" only meets some of the criteria. The research team applied the scoring tool to the testbeds identified in Section 3.4. We summarize the results of the evaluation in Section 4.2.

Table 8: Testbed Evaluation Tool.

| Evaluation Criteria | Weighting Factor | Solution 1 | | Solution 2 | |
|---|---|---|---|---|---|
| | | Rate | Score | Rate | Score |
| Data Collection & Performance Measures | 3 | 10 | 30 | 10 | 30 |
| Implementation Factors | 3 | 10 | 30 | 5 | 15 |
| Teaming Factors | 2 | 10 | 20 | 10 | 20 |
| Task Features | 2 | 10 | 20 | 0 | 0 |
| Scenario Authoring | 2 | 10 | 20 | 0 | 0 |
| Data Processing | 1 | 10 | 10 | 6 | 6 |
| System Architecture | 1 | 10 | 10 | 3 | 3 |
| Agents | 1 | 10 | 10 | 10 | 10 |
| Totals | | | 150 | | 84 |

## 4.2 Results

Two researchers from the project team rated each of the testbeds selected for evaluation. After completing their evaluation, they met to discuss any ratings that exceeded a difference of two points. The purpose of these discussions was to identify cases where evaluators had genuine disagreements on the particular evaluation quality, as opposed to times where the evaluation criteria may not have been applied the same way.

Table 9 summarizes the final scores for the evaluated testbeds. Additional details regarding the evaluation process and the results are provided in Appendix A.

Table 9: Consolidated Evaluation Results

| Evaluation Criteria | Weight | Evaluator 1 Rating | Score | Evaluator 2 Rating | Score |
|---|---|---|---|---|---|
| **Blocks World for Teams (BW4T)** | | | | | |
| Data Collection/ Performance Measures | 3 | 4 | 12 | 3 | 9 |
| Implementation Factors | 3 | 8 | 24 | 7 | 21 |
| Teaming Factors | 2 | 4 | 8 | 6 | 12 |

| Evaluation Criteria | Weight | Evaluator 1 Rating | Score | Evaluator 2 Rating | Score |
|---|---|---|---|---|---|
| Scenario Authoring | 2 | 5 | 10 | 5 | 10 |
| Task Features | 2 | 6 | 12 | 5 | 10 |
| Data Processing | 1 | 5 | 5 | 3 | 3 |
| System Architecture | 1 | 6 | 6 | 5 | 5 |
| Agents | 1 | 8 | 8 | 8 | 8 |
| Total Score | | | 85/150 | | 78/150 |
| **Black Horizon** | | | | | |
| Data Collection/ Performance Measures | 3 | 7 | 21 | 5 | 15 |
| Implementation Factors | 3 | 6 | 18 | 8 | 24 |
| Teaming Factors | 2 | 1 | 2 | 2 | 4 |
| Scenario Authoring | 2 | 6 | 12 | 8 | 16 |
| Task Features | 2 | 7 | 14 | 8 | 16 |
| Data Processing | 1 | 1 | 1 | 2 | 2 |
| System Architecture | 1 | 6 | 6 | 6 | 6 |
| Agents | 1 | 1 | 1 | 2 | 2 |
| Total Score | | | 75/150 | | 85/150 |
| *Overcooked! 2* | | | | | |
| Data Collection/ Performance Measures | 3 | 2 | 6 | 2 | 6 |
| Implementation Factors | 3 | 1 | 3 | 0 | 0 |
| Teaming Factors | 2 | 5 | 10 | 5 | 10 |
| Scenario Authoring | 2 | 0 | 0 | 0 | 0 |
| Task Features | 2 | 5 | 10 | 4 | 8 |
| Data Processing | 1 | 3 | 3 | 0 | 0 |
| System Architecture | 1 | 2 | 2 | 2 | 2 |
| Agents | 1 | 2 | 2 | 1 | 1 |
| Total Score | | | 36/150 | | 27/150 |
| *Overcooked!*-based "Gather Ingredients" Testbed Variation (Song et al., 2019) | | | | | |
| Data Collection/ Performance Measures | 3 | 1 | 3 | 0 | 0 |

| Evaluation Criteria | Weight | Evaluator 1 Rating | Score | Evaluator 2 Rating | Score |
|---|---|---|---|---|---|
| Implementation Factors | 3 | 5 | 15 | 5 | 15 |
| Teaming Factors | 2 | 0 | 0 | 0 | 0 |
| Scenario Authoring | 2 | 4 | 8 | 5 | 10 |
| Task Features | 2 | 0 | 0 | 0 | 0 |
| Data Processing | 1 | 1 | 1 | 1 | 1 |
| System Architecture | 1 | 4 | 4 | 5 | 5 |
| Agents | 1 | 4 | 4 | 5 | 5 |
| Total Score | | | 35/150 | | 36/150 |
| *Overcooked!*-based "Apple Juice" Testbed Variation (Gao et al., 2020) | | | | | |
| Data Collection/ Performance Measures | 3 | 1 | 3 | 2 | 6 |
| Implementation Factors | 3 | 6 | 18 | 6 | 18 |
| Teaming Factors | 2 | 3 | 6 | 3 | 6 |
| Scenario Authoring | 2 | 1 | 2 | 1 | 2 |
| Task Features | 2 | 4 | 8 | 2 | 4 |
| Data Processing | 1 | 5 | 5 | 5 | 5 |
| System Architecture | 1 | 4 | 4 | 5 | 5 |
| Agents | 1 | 5 | 5 | 6 | 6 |
| Total Score | | | 51/150 | | 52/150 |
| *Overcooked!*-based "Onion Soup" Testbed Variation (Carroll et al., 2019) | | | | | |
| Data Collection/ Performance Measures | 3 | 3 | 9 | 2 | 6 |
| Implementation Factors | 3 | 5 | 15 | 5 | 15 |
| Teaming Factors | 2 | 1 | 2 | 1 | 2 |
| Scenario Authoring | 2 | 4 | 8 | 5 | 10 |
| Task Features | 2 | 4 | 8 | 5 | 10 |
| Data Processing | 1 | 5 | 5 | 5 | 5 |
| System Architecture | 1 | 4 | 4 | 5 | 5 |
| Agents | 1 | 8 | 8 | 6 | 6 |
| Total Score | | | 51/150 | | 49/150 |

| Evaluation Criteria | Weight | Evaluator 1 Rating | Score | Evaluator 2 Rating | Score |
|---|---|---|---|---|---|
| *Overcooked!*-based "Sushi" Testbed Variation (Buehler et al., 2021) | | | | | |
| Data Collection/ Performance Measures | 3 | 7 | 21 | 2 | 6 |
| Implementation Factors | 3 | 0 | 0 | 0 | 0 |
| Teaming Factors | 2 | 1 | 2 | 1 | 2 |
| Scenario Authoring | 2 | 4 | 8 | 5 | 10 |
| Task Features | 2 | 6 | 12 | 5 | 10 |
| Data Processing | 1 | 5 | 5 | 5 | 5 |
| System Architecture | 1 | 0 | 0 | 1 | 1 |
| Agents | 1 | 0 | 0 | 1 | 1 |
| Total Score | | | 48/140 | | 35/140 |
| *Overcooked!*-based "Salad" Testbed Variation (Wu et al., 2021) | | | | | |
| Data Collection/ Performance Measures | 3 | 5 | 15 | 3 | 9 |
| Implementation Factors | 3 | 3 | 9 | 5 | 15 |
| Teaming Factors | 2 | 0 | 0 | 3 | 6 |
| Scenario Authoring | 2 | 5 | 10 | 5 | 10 |
| Task Features | 2 | 5 | 10 | 5 | 10 |
| Data Processing | 1 | 5 | 5 | 5 | 5 |
| System Architecture | 1 | 5 | 5 | 5 | 5 |
| Agents | 1 | 8 | 8 | 6 | 6 |
| Total Score | | | 62/150 | | 66/150 |
| *Neverwinter Nights* (SABRE Testbed; Leung et al., 2005) | | | | | |
| Data Collection/ Performance Measures | 3 | 6 | 18 | 8 | 24 |
| Implementation Factors | 3 | 6 | 18 | 6 | 18 |
| Teaming Factors | 2 | 6 | 12 | 6 | 12 |
| Scenario Authoring | 2 | 7 | 14 | 9 | 18 |
| Task Features | 2 | 9 | 18 | 8 | 16 |
| Data Processing | 1 | 5 | 5 | 5 | 5 |
| System Architecture | 1 | 7 | 7 | 7 | 7 |

| Evaluation Criteria | Weight | Evaluator 1 Rating | Score | Evaluator 2 Rating | Score |
|---|---|---|---|---|---|
| Agents | 1 | 2 | 2 | 6 | 6 |
| Total Score | | | 94/150 | | 106/150 |
| **ASIST Saturn+ (Huang et al., 2022a)** | | | | | |
| Data Collection/ Performance Measures | 3 | 8 | 24 | 10 | 30 |
| Implementation Factors | 3 | 6 | 18 | 7 | 21 |
| Teaming Factors | 2 | 4 | 8 | 4 | 8 |
| Scenario Authoring | 2 | 3 | 6 | 2 | 4 |
| Task Features | 2 | 7 | 14 | 6 | 12 |
| Data Processing | 1 | 7 | 7 | 8 | 8 |
| System Architecture | 1 | 5 | 5 | 7 | 7 |
| Agents | 1 | 5 | 5 | 5 | 5 |
| Total Score | | | 87/150 | | 95/150 |
| **ASIST Dragon** | | | | | |
| Data Collection/ Performance Measures | 3 | 9 | 27 | 10 | 30 |
| Implementation Factors | 3 | 8 | 24 | 7 | 21 |
| Teaming Factors | 2 | 3 | 6 | 3 | 6 |
| Scenario Authoring | 2 | 6 | 12 | 6 | 12 |
| Task Features | 2 | 7 | 14 | 6 | 12 |
| Data Processing | 1 | 7 | 7 | 8 | 8 |
| System Architecture | 1 | 6 | 6 | 7 | 7 |
| Agents | 1 | 5 | 5 | 5 | 5 |
| Total Score | | | 101/150 | | 101/150 |

Table 10 summarizes the final scores for each testbed.

Table 10: Consolidated Evaluation Scores

| Testbed | Evaluator 1 | Evaluator 2 |
|---|---|---|
| ASIST Dragon | 101 | 101 |
| *Neverwinter Nights* | 94 | 106 |
| ASIST Saturn+ | 87 | 95 |
| Blocks World for Teams (Johnson et al., 2009) | 85 | 78 |
| Black Horizon | 75 | 85 |
| *Overcooked!*-based "Salad" testbed variation (Wu et al., 2021) | 62 | 66 |
| *Overcooked!*-based "Onion Soup" testbed variation (Carroll et al., 2019) | 59 | 59 |
| *Overcooked!*-based "Apple Juice" testbed variation (Gao et al., 2020) | 51 | 52 |
| *Overcooked!*-based "Sushi" testbed variation (Buehler et al., 2021) | 48 | 35 |
| *Overcooked!*–based "Gather Ingredients" testbed variation (Song et al., 2019) | 35 | 36 |
| *Overcooked! 2* | 36 | 27 |

The highest scoring testbeds evaluated using our criteria were *Neverwinter Nights*, the two ASIST testbeds (Saturn+ and Dragon), BW4T, and Black Horizon. Our next steps in analyzing those testbed options are outlined in the next section.

## 5   NEXT STEPS

ASIST Dragon, ASIST Saturn+, BW4T, Black Horizon, and *Neverwinter Nights* all scored significantly higher than the other testbeds evaluated for implementation in the Human-AI teaming testbed. As a result, our future efforts will focus on these options.

The research team opted to deprioritize Black Horizon from further consideration. While Black Horizon supports multi-player operations, those players compete individually and the game does not support team-based operations. Eliminating Black Horizon leaves the team with three options for deeper consideration.

First, the team will continue to explore the viability of *Neverwinter Nights* as the basis of a testbed. One concern that warrants further consideration is the use of a commercial game as the core of the testbed. The researchers who responded to our survey showed a clear preference for an open source solution. We also want to develop a better sense for the nature of teamwork that the game can support and whether the game environment is capable of supporting autonomous agents as a part of gameplay. To explore these issues, we will continue to engage the research community in discussions regarding the benefits and drawbacks to using a commercial game as a foundation for the testbed. At the same time, we will work to gain firsthand experience with the game and the teaming experiences it can provide.

Second, we will continue to explore the viability of the ASIST testbeds. Like SABRE, which was based on *Neverwinter Nights*, the ASIST testbeds are built around Microsoft's *Minecraft* product. As a result, it might violate the open source preference expressed by our researcher respondents. The DARPA testbeds were also designed to support research in a different, but related, domain. The research team must determine the ease or difficulty with which we can adapt the testbeds to our needs.

Third, we will explore opportunities for applying the BW4T work. While BW4T is open source, it has drawbacks that give us pause. Significantly, much of the development of BW4T took place outside the United States, and it primary developers have stopped supporting this environment. Both of these factors may be incompatible with CMMC certification. To address these limitations, we will explore the development of a novel testbed that draws inspiration from BW4T (and the ASIST testbeds). We will want to determine:

1. Whether we can produce a testbed Concept of Operations that would score well using our testbed assessment rubric.
2. Whether we can produce that testbed cost-effectively enough to do so via an internal initiative.
3. Whether we can make that environment available to others via an open-source license.

This Page Intentionally Blank

Appendix A: Testbed Evaluation Report

Process Lead: Lillian Asiala

Decision Authority: Lillian Asiala

**Define the Issue.** The primary goal of Sonalysts' MUM-T initiative is to define a Synthetic Task Environment (STE) for conducting original Human-AI teaming research and working with other researchers (academics and Government laboratories). The original approach in the research plan was to define and develop a STE for this initiative. At the beginning of the design process, we conducted an outreach effort with researchers and Joint All-Domain Command and Control (JADC2) Subject Matter Experts (SMEs) to discover their priorities for an ideal STE. During that outreach effort, members of the research community mentioned that several open-source tools were available for use as a STE, and thus it was unnecessary to develop one. If a suitable tool already exists that meets our needs, then we can avoid the cost of developing a STE for this purpose and use an existing tool (or adapt/extend the tool) to conduct our research. The result of this Decision Analysis and Resolution (DAR) process will reveal whether we should proceed with the plan to develop a STE or adopt an existing tool.

**Establish the Evaluation Criteria.** The evaluation criteria for this DAR are based on a combination of our priorities for this project and the themes that emerged from surveying JADC2 SMEs and members of the research stakeholder community who conduct studies on Human-AI teaming. In the surveys, we asked respondents to indicate the degree to which they agreed or disagreed with a collection of STE features that we hypothesized would be of value to the research community. In addition to these Likert items, we also asked open-ended questions prompting respondents to offer suggestions about additional STE features.

We used these survey results to define testbed evaluation criteria. We conducted a thematic analysis of responses to open-ended questions to identify recurring ideas across answers, and then consolidated and combined these themes with ratings for related Likert items to form a master list of STE features. The entries on this master list were then grouped and sorted into a set of superordinate features. To narrow this list further, we dropped any features that had fewer than three associated open-ended responses or Likert-style questions. Additionally, we combined a few categories that were very similar in concept, and added several concepts that were not indicated in the survey data but were important to our efforts. This process produced a list of 23 features. As 23 is still too many features to use for testbed evaluation, the features in that list were further categorized into eight broad evaluation criteria categories, and features were used to create criteria descriptions.

We also developed a set of scoring standards that could be used to evaluate these criteria categories as they apply to various testbeds. Descriptions of these standards are documented in Table 11.

Table 11: Testbed Scoring Standards

| EVALUATION CRITERIA | SCORING STANDARD | | |
| --- | --- | --- | --- |
| | 0-3 | 4-6 | 7-10 |
| DATA COLLECTION /PERFORMANCE MEASURES | The testbed collects a few fixed data points. | The testbed collects a variety of outcome and process measures | The testbed includes rich instrumentation and likely supports the ability to define unique measures of interest. |
| IMPLEMENTATION FACTORS | The software architecture and/or problem space is complex. This will lead to long/costly development and/or modification timelines. | The architecture is reasonable, but various factors suggest that a good deal of software engineering and/or subject-matter expertise will be needed to develop/modify specific scenarios. | The system has a well-considered modular architecture. The problem space and/or authoring tools are simple enough to enable speedy and cost-effective development/modification. |
| TEAMING FACTORS | The size the team is small and fixed. Roles are not differentiated. Autonomous agents are not incorporated. Communication between teammates is not recorded. | Team size is appropriate (e.g., 4-6) and autonomous and human teammates can be assigned to multiple roles. Team communication is recorded. | Team size and configuration (autonomous/human teammates) is very flexible and roles are (or can be) clearly defined. All modes of communication are recorded for evaluation. |
| SCENARIO AUTHORING | Scenarios are fixed or only slightly adjustable. | There is a moderate level of author-ability in the scenarios, and the process seems reasonably straightforward. | The testbed includes scenario authoring tools that allow for a great deal of expressiveness and tailoring. |
| TASK FEATURES | The testbed tasks are fixed, artificial, and do not require a useful level of information processing. | The tasks require at least a little information processing and some level of customization is possible to modulate basic difficulty factors. | The testbed includes tasks that demand a great deal of information processing. The tasks include a large number of features that make it reasonable to modulate important complexity dimensions. |

| EVALUATION CRITERIA | SCORING STANDARD | | |
|---|---|---|---|
| | 0-3 | 4-6 | 7-10 |
| DATA PROCESSING | Data export is cumbersome and/or not possible. Data visualization is not possible. | Data can be exported in conventional formats (e.g., Excel or CSV). Analysis and visualization tools are not available. | There is significant support for data export in a variety of formats. Statistical analysis tools are provided. It is possible to generate data visualizations. |
| SYSTEM ARCHITECTURE | The architecture is fixed and monolithic. | The system architecture is reasonable and should not present insurmountable impediments to system modification. | The testbed uses a well-structured modular architecture. There are SDKs and/or other tools to support system maintenance/modification. |
| AGENTS | Agents cannot be incorporated into the system. The only option is "Wizard of Oz" emulation of agents. | Agents can be developed and/or interfaced with the testbed. The interface may require significant software engineering skill. | Well-defined APIs exist to facilitate the integration of agents. |

**Identify Alternative Solutions.** We reached out to a number of members of the research community to gather their recommendations on current publically available open-source testbeds to evaluate. Summaries of those responses are listed in Table 12.

Table 12: Summary of Stakeholder Feedback on the Availability of Open-Source Testbed

| Stakeholder | Input |
|---|---|
| Dr. Joan Johnston | Dr. Johnston provided recommendations for other researchers that we should contact (Dr. Katherine Cox, Dr. Melissa Walwanis, and Alan Lemon). |
| Dr. David Grimm | Dr. Grimm provided us with very useful references to Blocks World for Teams (BW4T). BW4T is an open-source testbed that has been used in some Human-AI teaming research. |
| Dr. Andrea Krausman | Dr. Krausman did not know of any open-source testbeds that would be of value. |
| Dr. Sarah Bibyk | Dr. Bibyk indicated that we should investigate the Defense Advanced Research Projects Agency (DARPA) Artificial Social Intelligence for Successful Teams (ASIST) project to determine if they would be a source of useful tools. |
| Dr. Peter Squire | Dr. Squire provided contact information for Marc Steinberg, an Office of Naval Research (ONR) researcher. |

| Stakeholder | Input |
|---|---|
| Dr. Marc Steinberg | Dr. Steinberg joined us in a very interesting discussion that led to the formation of the testbed taxonomy described in Technical Report #4. As part of those discussions, he noted that there were many "low end" testbeds, but few would support quality research in this domain. |
| Dr. Katherine Cox | Dr. Cox provided contact information for other researchers who might be able to help. |
| Dr. Melissa Walwanis (NAWCTSD) | Dr. Walwanis provided contact information for other researchers who might be able to help. |
| Mr. Alan Lemon | Mr. Lemon did not know of any suitable open-source testbeds. |

Based on this input, the research team decided to conduct a literature review to identify possible testbeds. We began with the BW4T system suggested by Dr. Grimm, and proceeded from there. In the next section, we document our evaluation of the systems that seemed suitable for our research.

**Select Evaluation Methods.** The research team evaluated each testbed with the criteria described in Table 11, using available information about the testbed found in published research articles. In some cases, further research was required to make an accurate assessment of whether testbed criteria were met or not.

Once evaluators completed their reviews, they discussed ratings that differed by more than two points on the scale. The purpose of these discussions was to determine whether the difference in rating was caused by the application of scoring standards, a misunderstanding of the literature, or a difference of opinion. In instances where scoring standards were not applied consistently, adjustments were made to ratings. In the case of confusion stemming from descriptions in the research literature, evaluators referenced the paper, further talked through the issue, and made adjustments accordingly. Differences of opinion resulted in no adjustment to the score.

Table 13: Summary of Evaluation Results

| Testbed | Evaluator 1 | Evaluator 2 |
|---|---|---|
| ASIST Dragon | 101 | 101 |
| *Neverwinter Nights* | 94 | 106 |
| ASIST Saturn+ | 87 | 95 |
| Blocks World for Teams (Johnson et al., 2009) | 85 | 78 |
| Black Horizon | 75 | 85 |
| *Overcooked!*-based "Salad" testbed variation (Wu et al., 2021) | 62 | 66 |
| *Overcooked!*-based "Onion Soup" testbed variation (Carroll et al., 2019) | 59 | 59 |
| *Overcooked!*-based "Apple Juice" testbed variation (Gao et al., 2020) | 51 | 52 |
| *Overcooked!*-based "Sushi" testbed variation (Buehler et al., 2021) | 48 | 35 |

| Testbed | Evaluator 1 | Evaluator 2 |
|---|---|---|
| *Overcooked!* –based "Gather Ingredients" testbed variation (Song et al., 2019) | 35 | 36 |
| *Overcooked! 2* | 36 | 27 |

**Selection**. BW4T, the ASIST testbeds, Black Horizon, and *Neverwinter Nights* all scored significantly higher than the other evaluated testbeds. With additional consideration, the research team decided to eliminate Black Horizon as a possible testbed because it is a competitive game focused on single player performance rather than a team-oriented mission environment.

The research team developed three possible courses of action based on the remaining highest-scoring testbeds.

First, the research team will purchase a license to play *Neverwinter Nights* and consider the benefits and drawbacks of using that commercial game as a foundation for the testbed. At this time, the research team needs to conduct a more in-depth investigation of the mechanics of the game and its authoring tools. Specifically, we need to evaluate the game's ability to support autonomous agents and teamwork more generally. One evident drawback of using *Neverwinter Nights* as the foundation for this testbed is that it is a commercial product, and the research stakeholders we consulted expressed a preference for an open-source solution.

This is also a concern with the foundation of the ASIST testbeds, Microsoft's Commercial Off-the-Shelf (COTS) game *Minecraft*. Nevertheless, we will also continue to explore the viability of the ASIST testbeds.

The research team decided to limit consideration of BW4T to the game's concept rather than the testbed itself. This is because BW4T is an open-source game primarily developed outside of the United States, and its developers have stopped supporting it. Consequently, is unclear whether it conforms to CMMC standards. We will develop a Concept of Operations (CONOPS) for software development, and assess the level of effort needed to pursue this course of action. Additionally, we will work with senior management to assess the feasibility of offering an open-source testbed.

Once these evaluations have been completed, and level of effort is assessed for both courses of action, a testbed will be selected and development will begin.

**Complete Testbed Evaluations**. The following tables and figures document the research team's evaluations of each of the testbeds.

Table 14: Blocks World for Teams (BW4T) Evaluation

| Evaluation Criteria | Weight | Evaluator 1 Rating | Score | Evaluator 2 Rating | Score |
|---|---|---|---|---|---|
| Data Collection/ Performance Measures | 3 | Rating: 4<br><br>Justification: BW4T creates a log file containing these data:<br>• Sequence: goal sequence (which block colors are to be dropped)<br>• Room: initial blocks per room<br>• Action: log of each action of a bot, with timestamp<br>• Total time: total time to complete task<br>• Agent summary: for each agent:<br>  o the bot type containing its handicaps<br>  o # correct drops in dropzone<br>  o # incorrect drops in dropzone<br>  o total time of standing still<br>  o # messages to other agents<br>  o # rooms entered | 12 | Rating: 3<br><br>Justification: The testbed collects a predefined set of values in a log file.<br><br>The system could be expanded to collect additional measures.<br><br>Therefore, I used the highest score in the first group. | 9 |

| Evaluation Criteria | Weight | Evaluator 1 Rating | Score | Evaluator 2 Rating | Score |
|---|---|---|---|---|---|
| Implementation Factors | 3 | Rating: 8<br><br>Justification: BW4T is an open-source environment with very simple game mechanics, which will make it easy to use, increasing speed of implementation and reducing development costs. | 24 | Rating: 7<br><br>Justification: The testbed is available via Github and written in Java, suggesting that it is not overly complex.<br><br>Other research teams have modified the base system for their research, suggesting that it is modifiable.<br><br>This led me to use the highest set of scores. I used the lower bounds because I haven't/can't review the code. | 21 |
| Teaming Factors | 2 | Rating: 4<br><br>Justification: BW4T supports teaming with multiple human and AI agents (2- any number). It does not have role differentiation. Communication is measured using a chat window, but the game captures no other communication modality. | 8 | Rating: 6<br><br>Justification: The literature suggests that BW4T supports multiple team members (at least 2, but usually more).<br><br>It seems like there is role differentiation (finders vs. movers), but not a lot.<br><br>This led me to use the upper bound of the middle category. | 12 |

| Evaluation Criteria | Weight | Evaluator 1 Rating | Score | Evaluator 2 Rating | Score |
|---|---|---|---|---|---|
| Scenario Authoring | 2 | Rating: 5<br><br>Justification: BW4T supports authoring via definition of:<br><br>- number of rooms<br>- room layout<br>- room contents<br>- nature of the target sequence<br><br>It does not have many job aids to simplify authoring. | 10 | Rating: 5<br><br>Justification: Authoring in this testbed includes defining rooms within a rectilinear play space, positioning blocks in rooms, and setting the collection order.<br><br>There does not seem to be a lot of expressiveness/tailoring involved.<br><br>Therefore, I selected the middle/middle value. | 10 |
| Task Features | 2 | Rating: 6<br><br>Justification: BW4T is an analog for search and rescue problems.<br><br>It is easy to learn/understand.<br><br>Features can be adjusted to make it more challenging (time available, number of block colors, rate of bot battery, information density in the environment, signal/noise ratio, uncertainty). It likely does not impose sufficiently complex information processing demands. | 12 | Rating: 5<br><br>Justification: The task involves some information process/situation assessment. Some level of modulation is possible by restricting the number of blocks relative to the required "drop sequence."<br><br>This led me to select the middle/middle value. | 10 |

| Evaluation Criteria | Weight | Evaluator 1 Rating | Score | Evaluator 2 Rating | Score |
|---|---|---|---|---|---|
| Data Processing | 1 | Rating: 5<br><br>Justification: BW4T creates a log file. | 5 | Rating: 3<br><br>Justification: Data are stored in a log file than must be manually extracted.<br><br>However, the log file has a known structure that should make the extraction reasonably direct.<br><br>The testbed does not include analysis or visualization support.<br><br>Therefore, I selected the upper bound of the lowest group. | 3 |
| System Architecture | 1 | Rating: 6<br><br>Justification: BW4T is an open source tool. It has a client-server architecture, which may allow it to support distributed research, with all labs having the same version of the software. A more technical analysis is required to assess whether it is sufficiently flexible/modular. | 6 | Rating: 5<br><br>Justification: I selected the middle/middle value because other teams have modified the base testbed, indicating that it is possible.<br><br>However, this is a dimension that would be better assessed by a software engineer. | 5 |

| Evaluation Criteria | Weight | Evaluator 1 Rating | Score | Evaluator 2 Rating | Score |
|---|---|---|---|---|---|
| Agents | 1 | Rating: 8<br><br>Justification: BW4T was designed with agent integration in mind. | 8 | Rating: 8<br><br>Justification: BW4T was designed with interfacing agents in mind. Tools exist to create agents and the system uses a well-defined interface.<br><br>Without a closer analysis of the code, it is difficult to assess how difficult it will be to use the interface to integrate third-party agents. | 8 |
| Total | | | 85/150 | | 78/150 |

The unweighted scores for each evaluation dimension are shown in Figure 12.

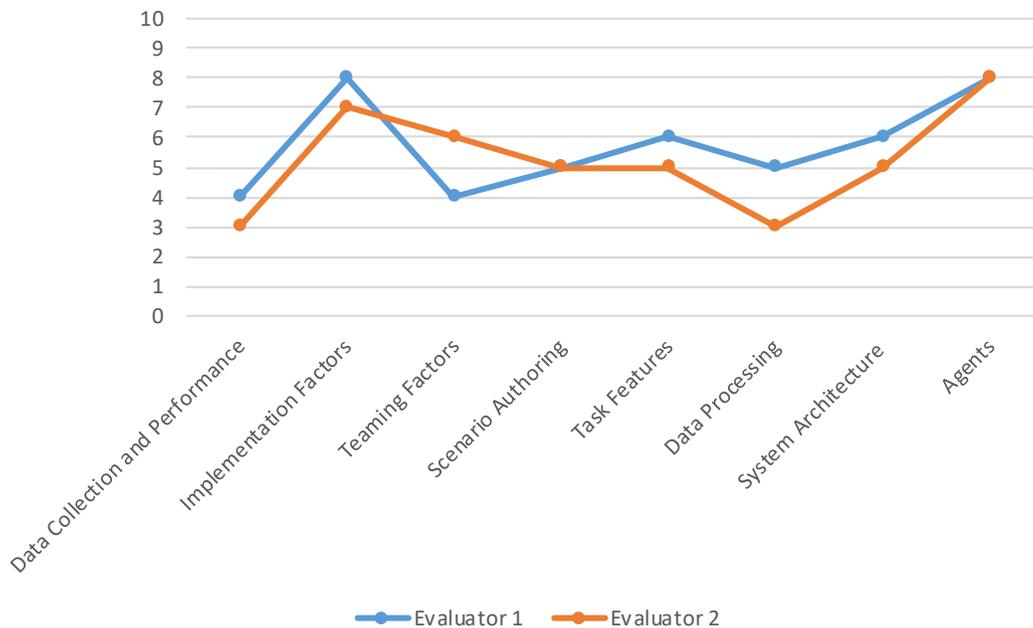

**Figure 12: Evaluation Dimension Scores for BW4T**

Table 15: Black Horizon Evaluation

| Evaluation Criteria | Weight | Evaluator 1 Rating | Score | Evaluator 2 Rating | Score |
|---|---|---|---|---|---|
| Data Collection/ Performance Measures | 3 | Rating: 7<br><br>Justification: Black Horizon appears to be well-instrumented, collecting the following variables:<br><br>• Overall mission result (succeeded/failed)<br>• Badges earned (awarded based on cumulative performance)<br>• Mission objectives and performance detail<br><br>Objectives were defined for each mission, and specific metrics were evaluated based on those objectives. Black Horizon recorded whether or not standards for these metrics were met. | 21 | Rating: 5<br><br>Justification: Black Horizon measures time, accuracy, and "strategic" goals. The current scoring is based on pass/fail criteria or a point-based schema. Data collection/performance assessment is tightly coupled to the goals of a given scenario. | 15 |
| Implementation Factors | 3 | Rating: 6<br><br>Justification: Sonalysts created Black Horizon and owns the components of the game, giving this a high probability of being easy and relatively fast to implement as a testbed. However, a reasonable amount of effort is needed to create and program test scenarios on the game engine. | 18 | Rating: 8<br><br>Justification: Black Horizon is based on a modular architecture that Sonalysts understands well. Authoring tools are available, but require expertise to use well. Black Horizon employs the Unity game engine which facilitates development. | 24 |

| Evaluation Criteria | Weight | Evaluator 1 Rating | Score | Evaluator 2 Rating | Score |
|---|---|---|---|---|---|
| Teaming Factors | 2 | Rating: 1<br><br>Justification: Black Horizon has a multi-player version, with mission sizes ranging from 2-4 players. However, players compete with one another, rather than filling complementary roles. Roles are not differentiated and autonomous agents are not incorporated. | 2 | Rating: 2<br><br>Justification: Black Horizon missions seem to be completed by individuals, not teams. There is a multiplayer capability, but the players seem to be competitive, rather than cooperative. Some platforms are controlled by intelligent agents. | 4 |
| Scenario Authoring | 2 | Rating: 6<br><br>Justification: A set of fixed scenarios come with the game. Creating new scenarios is possible given the system architecture, but would take effort and expertise to create. | 12 | Rating: 8<br><br>Justification: Black Horizon currently uses a drag-and-drop system which does require some expertise. | 16 |
| Task Features | 2 | Rating: 7<br><br>Justification: The task environment for Black Horizon is designed to train operators, and therefore has high fidelity for training the skills in question. Task difficulty is modulated through levels of difficulty, and is not necessarily modular within a single mission. | 14 | Rating: 8<br><br>Justification: Black Horizon seems to require a good deal of information processes and it has a well-defined complexity progression, suggesting that difficulty is controllable. | 16 |

| Evaluation Criteria | Weight | Evaluator 1 Rating | Score | Evaluator 2 Rating | Score |
|---|---|---|---|---|---|
| Data Processing | 1 | Rating: 1<br><br>Justification: It is not clear what data export options there are with Black Horizon, none are described. | 1 | Rating: 2<br><br>Justification: Black Horizon does not currently support data export, but it would not be hard to add that capability. There is no on-board analysis/visualization. | 2 |
| System Architecture | 1 | Rating: 6<br><br>Justification: Black Horizon was developed using Unity. This game engine offers a software framework to build games and simulations. It separates game function from specifics, so that different details could be plugged into general game components, indicating it should be relatively easy to make modifications, but tools do not necessarily exist to easily do so. | 6 | Rating: 6<br><br>Justification: Black Horizon is based on a modular architecture that Sonalysts understands well. There are no readily apparent SDKs/APIs. | 6 |

| Evaluation Criteria | Weight | Evaluator 1 Rating | Score | Evaluator 2 Rating | Score |
|---|---|---|---|---|---|
| Agents | 1 | Rating: 1<br><br>Justification: Agents were not used in Black Horizon, but Wizard of Oz techniques could be used to examine autonomous agent dynamics with human teammates. However, this issue is compounded by the fact that this isn't much of a "teaming" environment, but rather a competitive one. | 1 | Rating: 2<br><br>Justification: There is little/no support for integrating agents as teammates in the current version of Black Horizon. | 2 |
| Total | | | 75/150 | | 85/150 |

The unweighted scores for each evaluation dimension are shown in Figure 13.

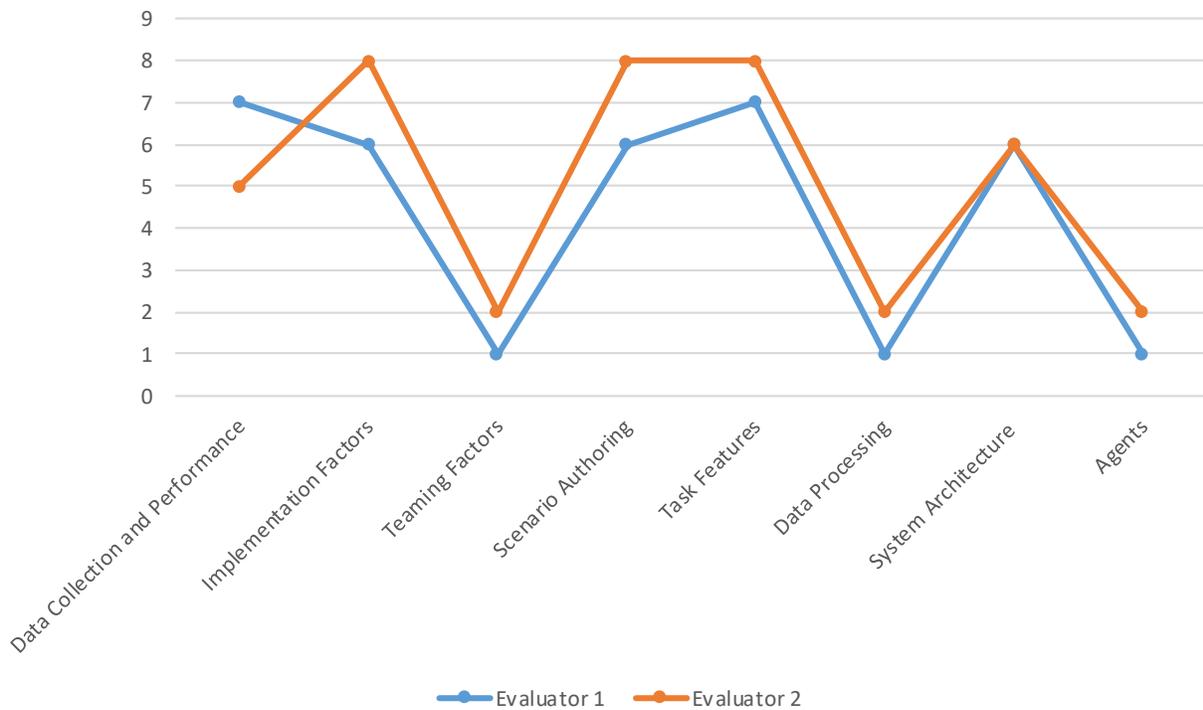

Figure 13: Evaluation Dimension Scores for Black Horizon

Table 16: *Overcooked! 2* Evaluation

| Evaluation Criteria | Weight | Evaluator 1 Rating | Score | Evaluator 2 Rating | Score |
|---|---|---|---|---|---|
| Data Collection/ Performance Measures | 3 | Rating: 2<br><br>Justification: The overall player score is based on correct recipes delivered in a particular timeframe. | 6 | Rating: 2<br><br>Justification: In the Bishop et al. work (2020), it appears that they only used the game score provided automatically. They also mention tracking "productive chef action." However, in reviewing game documentation, I can't find any reference to that score, so perhaps it was manually counted. Overcooked! 2 collects fewer datapoints than BW4T, so I assigned a lower score. | 6 |
| Implementation Factors | 3 | Rating: 1<br><br>Justification: Implementing Overcooked! 2 in a testbed will require purchasing the game, which is associated with cost, but implementation would be fast and low cost. However, implementation would be impacted negatively by the lack of detailed data collection and scenario authoring. | 3 | Rating: 0<br><br>Justification: Modification of Overcooked! 2 is not possible. I saw no mentions of authoring tools. | 0 |

| Evaluation Criteria | Weight | Evaluator 1 Rating | Score | Evaluator 2 Rating | Score |
|---|---|---|---|---|---|
| Teaming Factors | 2 | Rating: 5<br><br>Justification: The game generally supports small teams (up to four players). Roles were not well-differentiated and players often have to switch their responsibilities based on the changing dynamics of the task. | 10 | Rating: 5<br><br>Justification: Overcooked! 2 supports up to 4 team members. Chat-based communication is supported. I did not read anything indicating that communications were stored. | 10 |
| Scenario Authoring | 2 | Rating: 0<br><br>Justification: Scenario definition did not appear to be possible with the COTS version of the game. | 0 | Rating: 0<br><br>Justification: I have not read anything suggesting that Overcooked! 2 supports scenario authoring. | 0 |

| Evaluation Criteria | Weight | Evaluator 1 Rating | Score | Evaluator 2 Rating | Score |
|---|---|---|---|---|---|
| Task Features | 2 | Rating: 5<br><br>Justification: The tasks presented in the COTS game lack complex information challenges and decision-making that many military tasks require. However, aspects of the task environment can be modified to modulate task difficulty, and enhance the need for push/pull forms of communication among teammates, and teammates are required to prioritize tasks and avoid distraction (Tossell et al., 2020). Other aspects of the game have been praised for emulating common team operation stressors, such as creating realistic levels of chaos and confusion. | 10 | Rating: 4<br><br>Justification: Overcooked! 2 does require teamwork (in multiplayer mode). However, my assessment is that it doesn't involve much information processing. The levels are increasingly difficult, but that's beyond the control of the research team. | 8 |
| Data Processing | 1 | Rating: 3<br><br>Justification: The research using this testbed does not make any mention of data visualization. Descriptions of data export were also missing, though it is assumed to be possible given data had to be collected and exported to be analyzed for studies. | 3 | Rating: 0<br><br>Justification: There is no indication that data export, analysis, or visualization are supported. | 0 |

| Evaluation Criteria | Weight | Evaluator 1 Rating | Score | Evaluator 2 Rating | Score |
|---|---|---|---|---|---|
| System Architecture | 1 | Rating: 2<br><br>Justification: Bishop et al. (2020) mentioned the system architecture for the game being amenable to virtual teams that are not co-located, through their description of the Wizard of Oz paradigm. In their study, two teammates played in different rooms, with the research confederate as one player posing as both human and autonomous agent in different conditions. This suggests that teams may be co-located.<br><br>The system architecture itself doesn't appear to allow for modification. | 2 | Rating: 2<br><br>Justification: The system is not an open architecture and cannot be modified. Distributed teams are possible. | 2 |
| Agents | 1 | Rating: 2<br><br>Justification: The base game does not use agents, but research articles used Wizard of Oz techniques where humans posed as AI agents to measure changes in human teammate perception. | 2 | Rating: 1<br><br>Justification: Agents are not supported. However, Wizard of Oz techniques can be used to simulate agents. | 1 |
| Total | | | 36/150 | | 27/150 |

The unweighted scores for each evaluation dimension are shown in Figure 14.

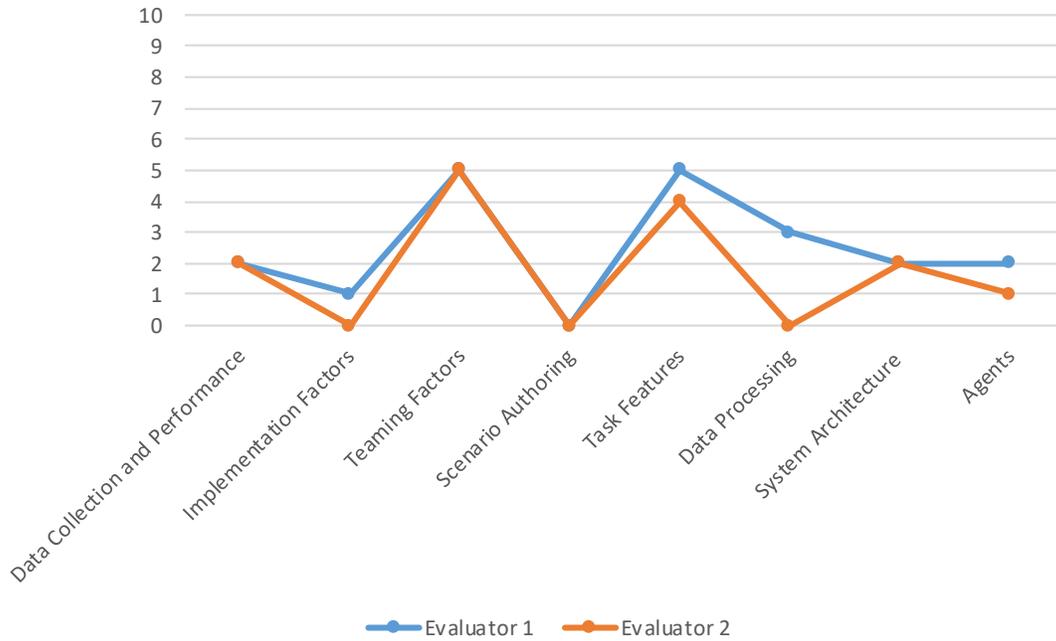

**Figure 14: Evaluation Dimension Scores for Overcooked! 2**

Table 17: *Overcooked!*-based "Gather Ingredients" Testbed Variation

| Evaluation Criteria | Weight | Evaluator 1 Rating | Score | Evaluator 2 Rating | Score |
|---|---|---|---|---|---|
| Data Collection/ Performance Measures | 3 | Rating: 1<br><br>Justification: This testbed collects overall score, but other variables were either not collected or were not described by Song et al. (2019). The researchers collected a "learning speed" score, but that did not appear to be obtained in the testbed. | 3 | Rating: 0<br><br>Justification: I could not ascertain what, if any, data the Song et al. (2019) testbed gathered. | 0 |

| Evaluation Criteria | Weight | Evaluator 1 Rating | Score | Evaluator 2 Rating | Score |
|---|---|---|---|---|---|
| Implementation Factors | 3 | Rating: 5<br><br>Justification: Code appears to be available, but more information is needed to assess this dimension. | 15 | Rating: 5<br><br>Justification: The task environment is simple and the code base is open source. This suggests that implementation should not be complex and time consuming. However, we may want to ask a software engineer to validate this. | 15 |
| Teaming Factors | 2 | Rating: 0<br><br>Justification: This testbed didn't appear to be designed for teaming. Only one agent performs at a time. | 0 | Rating: 0<br><br>Justification: The testbed only supports a single user. That user is an agent. | 0 |
| Scenario Authoring | 2 | Rating: 4<br><br>Justification: The paper describes nine baselines with four "perspectives", which are programmed into the testbed. However, it is not clear how straightforward it is to create new scenarios. | 8 | Rating: 5<br><br>Justification: Song et al. (2019) created multiple scenarios by defining various ingredient orders. No authoring tools were described. | 10 |

| Evaluation Criteria | Weight | Evaluator 1 Rating | Score | Evaluator 2 Rating | Score |
|---|---|---|---|---|---|
| Task Features | 2 | Rating: 0<br><br>Justification: The goal of Overcooked! is to get an agent to fetch multiple ingredients in a special sequence in accordance with a pick list (this list changes by episode). This task didn't appear to require team coordination whatsoever. | 0 | Rating: 0<br><br>Justification: The task did not require coordination. | 0 |
| Data Processing | 1 | Rating: 1<br><br>Justification: Unclear whether data is easily exported from the testbed. No data visualization capabilities were described. | 1 | Rating: 1<br><br>Justification: The nature of the Song et al. (2019) work suggests that it is possible to export captured data. No visualization tools seem to exist. | 1 |
| System Architecture | 1 | Rating: 4<br><br>Justification: The system is open-source but it is unclear how easy it would be to modify based on the description. | 4 | Rating: 5<br><br>Justification: The system is open source. It was not possible to determine the quality of its architecture, but it should not present insurmountable challenges. There do not appear to be any SDKs or other tools to support system maintenance/modification. | 5 |

| Evaluation Criteria | Weight | Evaluator 1 Rating | Score | Evaluator 2 Rating | Score |
|---|---|---|---|---|---|
| Agents | 1 | Rating: 4<br><br>Justification: A single agent is described as being designed for this testbed, but it is unclear how easy it would be to "plug in" others. This might take significant software engineering to accomplish. | 4 | Rating: 5<br><br>Justification: Song et al. (2019) developed agents and embedded them within the game. There was no mention of doing this via APIs. | 5 |
| Total | | | 35/150 | | 36/150 |

The unweighted scores for each evaluation dimension are shown in Figure 15.

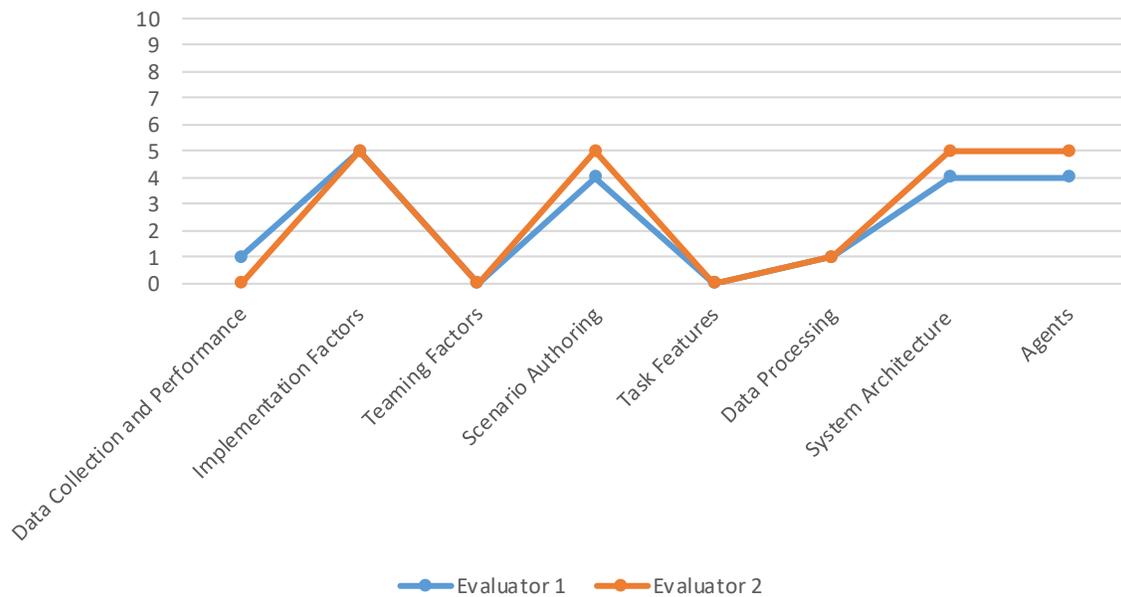

**Figure 15: Evaluation Dimension Scores for Overcooked!-based "Gather Ingredients" Testbed Variation**

Table 18: *Overcooked!*-based "Apple Juice" Testbed Variation Evaluation

| Evaluation Criteria | Weight | Evaluator 1 Rating | Score | Evaluator 2 Rating | Score |
|---|---|---|---|---|---|
| Data Collection/ Performance Measures | 3 | Rating: 1<br><br>Justification: This testbed does not seem to capture much data on the task, other than task timing. | 3 | Rating: 2<br><br>Justification: Gao et al. (2020) collected one metric – the time required to complete each order. | 6 |
| Implementation Factors | 3 | Rating: 6<br><br>Justification: This testbed appeared to be open-source, making implementation relatively straightforward. | 18 | Rating: 6<br><br>Justification: The task environment is simple and the code base is open source and based on the Unreal engine. This suggests that implementation should not be complex and time consuming. However, we may want to ask a software engineer to validate this. | 18 |
| Teaming Factors | 2 | Rating: 3<br><br>Justification: The size of teams in this testbed seemed to be restricted to dyads of two: one human and one agent. The agent also seemed to have a singular role assisting the human, instead of being able to fill a number of roles flexibly. Communication within the dyad was both manipulated and measured. | 6 | Rating: 3<br><br>Justification: In Gao et al. (2020) only two team members were involved (one human and one agent). It appears that the agent and human had differentiated roles. | 6 |

| Evaluation Criteria | Weight | Evaluator 1 Rating | Score | Evaluator 2 Rating | Score |
|---|---|---|---|---|---|
| Scenario Authoring | 2 | Rating: 1<br><br>Justification: Goals and sub-goals of the task are assigned to the human or the autonomous agent. It is unclear from the author's description how amenable the testbed is to task/goal authoring within the main "juice-making" task, but the description appears to be fairly constrained. | 2 | Rating: 1<br><br>Justification: In Gao et al. (2020), it seems that only one kitchen was used and the level of difficulty, interdependence, time pressure, etc. was not varied. | 2 |
| Task Features | 2 | Rating: 4<br><br>Justification: The authors do not mention ability to modulate task difficulty, but instead are focused on modulating the information communicated from the autonomous agent to the human. Psychological fidelity with military operations or complex human/autonomous teams more generally is dubious in this testbed. | 8 | Rating: 2<br><br>Justification: The required level of coordination appears to be fixed. The task requires little information processing and decision-making. | 4 |

| Evaluation Criteria | Weight | Evaluator 1 Rating | Score | Evaluator 2 Rating | Score |
|---|---|---|---|---|---|
| Data Processing | 1 | Rating: 5<br><br>Justification: The article does not detail how data is processed by the testbed. Data was assumed to be exported for processing and analysis. | 5 | Rating: 5<br><br>Justification: The nature of the Gao et al. (2020) work suggests that it is possible to export captured data. No visualization tools seem to exist. | 5 |
| System Architecture | 1 | Rating: 4<br><br>Justification: This testbed is open-source. The researchers did not describe the flexibility of the architecture for this testbed. It was created to mimic "Overcooked! 2" using "unreal" engine. More information via follow-up with authors is required. | 4 | Rating: 5<br><br>Justification: The system is open source. It was not possible to determine the quality of its architecture, but it should not present insurmountable challenges. There do not appear to be any SDKs or other tools to support system maintenance/modification. | 5 |
| Agents | 1 | Rating: 5<br><br>Justification: One agent of limited ability was incorporated into this testbed, indicating agents can be developed and incorporated. | 5 | Rating: 6<br><br>Justification: Gao et al. (2020) developed agents and embedded them within the game. There was no mention of doing this via APIs. | 6 |
| Total | | | 51/150 | | 52/150 |

The unweighted scores for each evaluation dimension are shown in Figure 16.

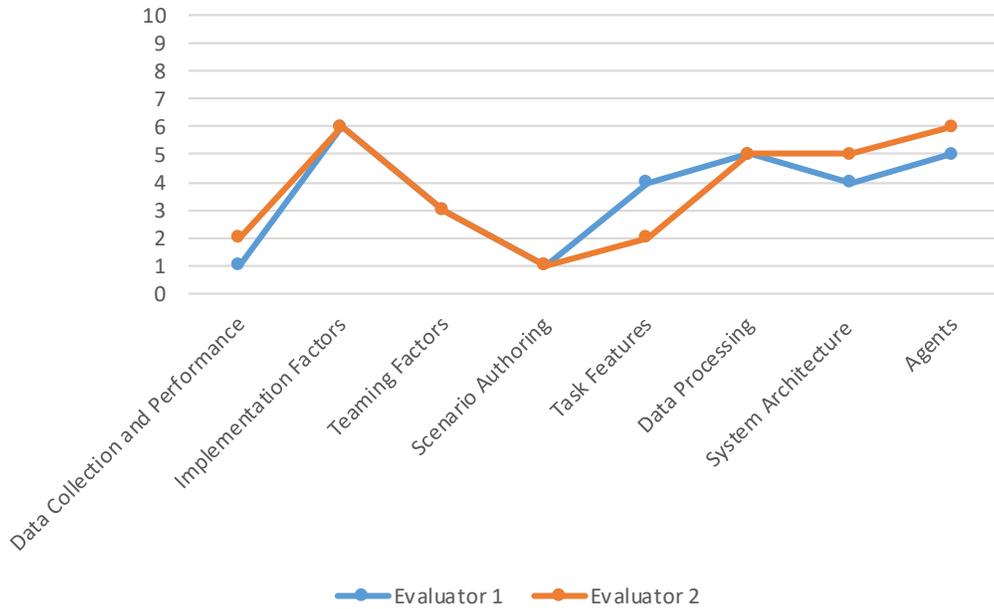

**Figure 16: Evaluation Dimension Scores for Overcooked!-based "Apple Juice" Testbed Variation**

Table 19: *Overcooked!*-based "Onion Soup" Testbed Variation Evaluation

| Evaluation Criteria | Weight | Evaluator 1 Rating | Score | Evaluator 2 Rating | Score |
|---|---|---|---|---|---|
| Data Collection/ Performance Measures | 3 | Rating: 3<br><br>Justification: Beyond the number of goal tasks that can be achieved in one play session, (cumulative "rewards") the description of this testbed did not contain more instrumentation that would facilitate the collection of other variables. This limits data collection to specified outcome variables. | 9 | Rating: 2<br><br>Justification: Carroll et al. (2019) refer to gathering data on "reward" points, but these are not defined. It may be possible to extend the testbed to collect additional data. | 6 |

| Evaluation Criteria | Weight | Evaluator 1 Rating | Score | Evaluator 2 Rating | Score |
|---|---|---|---|---|---|
| Implementation Factors | 3 | Rating: 5<br><br>Justification: The code-base for this testbed is open-source, meaning implementation would not be difficult. | 15 | Rating: 5<br><br>Justification: The task environment is simple and the code base is open source. This suggests that implementation should not be complex and time consuming. However, we may want to ask a software engineer to validate this. | 15 |
| Teaming Factors | 2 | Rating: 1<br><br>Justification: This testbed had small teams (single human/agent dyads), and was described as having inflexible roles based on where the human and agent were located in the kitchen environment (although there may be flexibility in the placement of human and agent in the kitchen layout). Communication between the agent and human was not collected as a variable, it is unclear if it was (or could be) recorded. | 2 | Rating: 1<br><br>Justification: The testbed only supports pairs of players and roles do not appear to be differentiated. | 2 |

| Evaluation Criteria | Weight | Evaluator 1 Rating | Score | Evaluator 2 Rating | Score |
|---|---|---|---|---|---|
| Scenario Authoring | 2 | Rating: 4<br><br>Justification: Scenario authoring in this version of the game is limited to features of the specific soup-making task the researchers developed for the study. Based on the description it is unclear how many aspects of the task can be changed to accommodate research questions. Coordination challenges through kitchen layout and agent training were two adjustable features based on the article description. | 8 | Rating: 5<br><br>Justification: Carroll et al. (2019) created four scenarios, indicating that authoring is possible. This is likely done in the code, but I was unable to confirm this. There was no mention of authoring tools. | 10 |
| Task Features | 2 | Rating: 4<br><br>Justification: This version of the game was able to modulate coordination challenges through kitchen layout, however other elements (such as uncertainty or signal-to-noise ratio) were not present and thus not capable of being modified. | 8 | Rating: 5<br><br>Justification: The required level of coordination appears to be adjustable. However, the task requires little information processing and decision-making. | 10 |

| Evaluation Criteria | Weight | Evaluator 1 Rating | Score | Evaluator 2 Rating | Score |
|---|---|---|---|---|---|
| Data Processing | 1 | Rating: 5<br><br>Justification: The description in the article did not lend itself to rating this feature. Data visualization was not mentioned at all, and likely was not a part of testbed design. Data export was likely possible due to subsequent analysis of the data. | 5 | Rating: 5<br><br>Justification: The nature of the Carroll et al. (2019) work suggests that it is possible to export captured data. No visualization tools seem to exist. | 5 |
| System Architecture | 1 | Rating: 4<br><br>Justification: The system is open-source. However, from the description it is unclear how easy it will be to access and use it for research. It was not used to link human teammates that were not co-located. | 4 | Rating: 5<br><br>Justification: The system is open source. It was not possible to determine the quality of its architecture, but it should not present insurmountable challenges. There do not appear to be any SDKs or other tools to support system maintenance/modification. | 5 |

| Evaluation Criteria | Weight | Evaluator 1 Rating | Score | Evaluator 2 Rating | Score |
|---|---|---|---|---|---|
| Agents | 1 | Rating: 8<br><br>Justification: The agents in this testbed were trained either with self-play, other autonomous agents, or using a human learning model. This is a good example of the ability to "plug in" agents with differing capabilities. However, agents in the study were limited to the human/agent dyad, so multiple agents of differing abilities did not work together. | 8 | Rating: 6<br><br>Justification: Carroll et al. (2019) developed agents and embedded them within the game. There was no mention of doing this via APIs. | 6 |
| Total |  |  | 51/150 |  | 49/150 |

The unweighted scores for each evaluation dimension are shown in Figure 17.

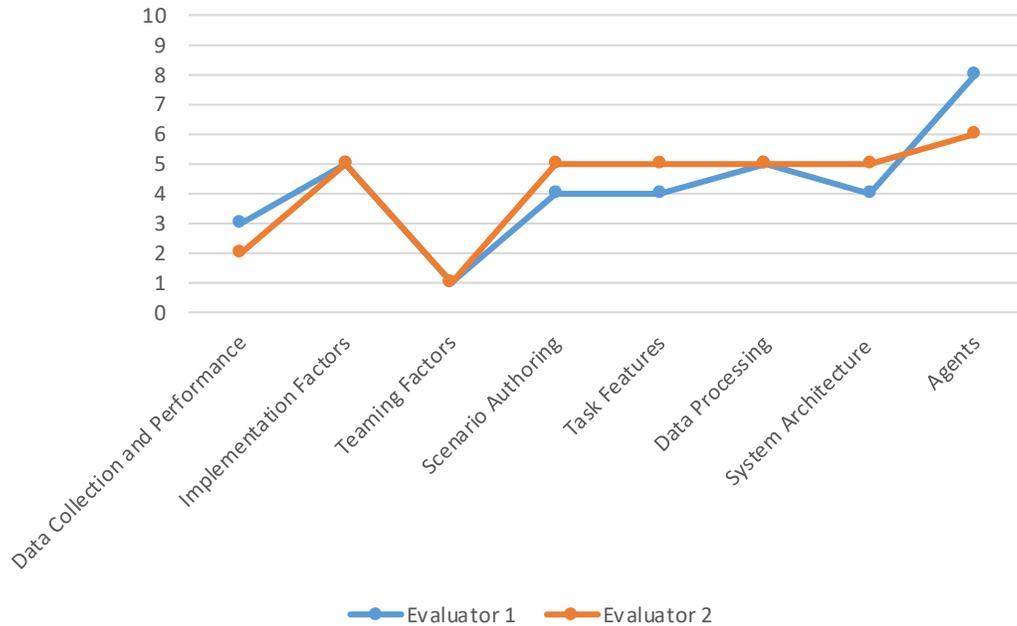

**Figure 17:** Evaluation Dimension Scores for Overcooked!-based "Onion Soup" Testbed Variation

**Table 20:** *Overcooked!*-based "Sushi" Testbed Variation Evaluation

| Evaluation Criteria | Weight | Evaluator 1 Rating | Score | Evaluator 2 Rating | Score |
|---|---|---|---|---|---|
| Data Collection/ Performance Measures | 3 | Rating: 7<br><br>Justification: The fully integrated testbed (*i.e.,* game software, interactive equipment, *etc*.) was instrumented to collect eye tracking from human participants, and track action duration, errors, and communication between human and robot. This version of the game demonstrated to collect a range of performance measures. | 21 | Rating: 2<br><br>Justification: Buehler, Adamy, & Weisswange (2021) indicated that they collected data on number of errors and the length of error sequences. They do not explicitly say whether their environment collected those data, but it seems to be a reasonable assumption. | 6 |

| Evaluation Criteria | Weight | Evaluator 1 Rating | Score | Evaluator 2 Rating | Score |
|---|---|---|---|---|---|
| Implementation Factors | 3 | Rating: 0<br><br>Justification: This testbed would be difficult to implement as a study environment, even if the game is made available be the researchers (and the code base is not currently available), the special equipment designed for the robot interface and the robot agent pose issues implementing this option. | 0 | Rating: 0<br><br>Justification: The task environment is simple. However, the code base is not available. | 0 |
| Teaming Factors | 2 | Rating: 1<br><br>Justification: Based on the description, communication with the robot agent appeared to be visual (showing a recipe or indicating something in the cooking space), instead of verbal or chat-based. This testbed focused on human/robot dyads instead of larger teams of 4-12, and the roles of the human and robot were fairly undifferentiated. | 2 | Rating: 1<br><br>Justification: The testbed only supports pairs of players and roles do not appear to be differentiated. | 2 |

| Evaluation Criteria | Weight | Evaluator 1 Rating | Score | Evaluator 2 Rating | Score |
|---|---|---|---|---|---|
| Scenario Authoring | 2 | Rating: 4<br><br>Justification: As with other version of this testbed, it is not described to be capable of creating multiple scenarios for developing specific tasks to meet research objectives. Rather, the task has been designed specifically to address the researcher's questions. | 8 | Rating: 5<br><br>Justification: Buehler, Adamy, & Weisswange (2021) defined six recipes. This is likely done in the code, but I was unable to confirm this. There was no mention of authoring tools. | 10 |

| Evaluation Criteria | Weight | Evaluator 1 Rating | Score | Evaluator 2 Rating | Score |
|---|---|---|---|---|---|
| Task Features | 2 | Rating: 6<br><br>Justification: The task developed for this testbed was a cooperative robot/human sushi making task, where the human and robot work together to process ingredients and assemble different sushi rolls cooperatively according to customer orders. The task intentionally presents challenges to:<br><br>- Human situation awareness<br>- Task planning<br>- Teammate coordination (including physical coordination)<br>- Task difficulty through multiple recipes generating in random order, and confusion with similar-looking ingredients<br><br>The task does not have close fidelity to a military task, but presents some team stressors that emulate the stressors of those environments. While the task presents these stressors, it is not clear to what extent those variables can be adjusted to different levels. | 12 | Rating: 5<br><br>Justification: The required level of coordination appears to be adjustable. However, the task requires little information processing and decision-making. | 10 |

| Evaluation Criteria | Weight | Evaluator 1 Rating | Score | Evaluator 2 Rating | Score |
|---|---|---|---|---|---|
| Data Processing | 1 | Rating: 5<br><br>Justification: Data processing was not explored in the article. Data visualization is unlikely, data export was likely due to analysis of data. | 5 | Rating: 5<br><br>Justification: The nature of the Buehler, Adamy, & Weisswange (2021) work suggests that it is possible to export captured data. No visualization tools seem to exist. | 5 |
| System Architecture | 1 | Rating: 0<br><br>Justification: Not open-source and not available. | 0 | Rating: 1<br><br>Justification: The system is not an open architecture and cannot be modified. | 1 |
| Agents | 1 | Rating: 0<br><br>Justification: The agent is a robot in this testbed, and reacts to behaviors of the human without explicit coordination happening. The ability to plug in different robots of differing capabilities was not explored. | 0 | Rating: 1<br><br>Justification: Buehler, Adamy, & Weisswange (2021) defined agents, but it appears that they were not integrated with the testbed. Instead, agent functionality was simulated via a Wizard of Oz paradigm. | 1 |
| Total | | | 48/150 | | 35/150 |

The unweighted scores for each evaluation dimension are shown in Figure 18.

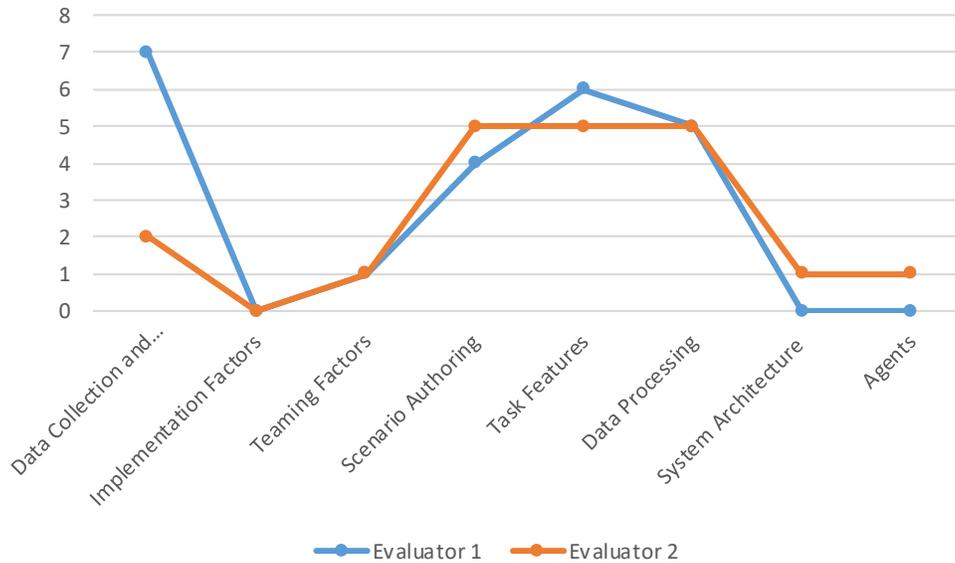

**Figure 18: Evaluation Dimension Scores for Overcooked!-based "Sushi" Testbed Variation**

**Table 21: *Overcooked!*-based "Salad" Testbed Variation Evaluation**

| Evaluation Criteria | Weight | Evaluator 1 Rating | Score | Evaluator 2 Rating | Score |
|---|---|---|---|---|---|
| Data Collection/ Performance Measures | 3 | Rating: 5<br><br>Justification: The testbed presents a multi-agent "object oriented" cooking task with problems that have a hierarchical organization of sub-tasks. The testbed was instrumented to measure agent timing and completion of subtasks, although the code for the testbed could likely be modified to measure other aspects of performance. | 15 | Rating: 3<br><br>Justification: Wu et al., (2021) collected data on three metrics:<br><br>- The number of time steps to complete the full recipe<br>- The total fraction of sub-tasks completed<br>- The average number of "shuffles" (*i.e.*, any action that negates the previous action) | 9 |

| Evaluation Criteria | Weight | Evaluator 1 Rating | Score | Evaluator 2 Rating | Score |
|---|---|---|---|---|---|
| Implementation Factors | 3 | Rating: 3<br><br>Justification: It is not clear whether this testbed would be available to modify for both human and agent use, to make it a HAT testbed. More information is needed for assessment. | 9 | Rating: 5<br><br>Justification: The task environment is simple and the code base is open source. This suggests that implementation should not be complex and time consuming. However, we may want to ask a software engineer to validate this. | 15 |
| Teaming Factors | 2 | Rating: 0<br><br>Justification: This testbed was not designed to facilitate humans and agents working together, but rather to look at how different approaches to agent design would impact performance. | 0 | Rating: 3<br><br>Justification: The testbed supports teams of two or three. However, in Wu et al. (2021) all players were agents. | 6 |
| Scenario Authoring | 2 | Rating: 5<br><br>Justification: Authors modified recipes, kitchen layout, and task difficulty. | 10 | Rating: 5<br><br>Justification: The Wu et al. (2021) game seems to support three or more kitchen configurations (no divider, partial divider, complete divider), multiple recipe variations. There was no mention of authoring tools. | 10 |

| Evaluation Criteria | Weight | Evaluator 1 Rating | Score | Evaluator 2 Rating | Score |
|---|---|---|---|---|---|
| Task Features | 2 | Rating: 5<br><br>Justification: The task presented in this testbed was to complete a recipe quickly, by processing and combining multiple ingredients. The testbed presented several kitchen configurations that posed certain performance challenges. While the task does not have fidelity with military tasks, it was designed to emulate certain teaming challenges to differently trained agents. These challenges were presented not only by the task environment but also in how agents were trained.<br><br>The aspects of the task that can be modulated require a closer look at the game code to adequately assess. | 10 | Rating: 5<br><br>Justification: The required level of coordination appears to be adjustable. However, the task requires little information processing and decision-making. | 10 |

| Evaluation Criteria | Weight | Evaluator 1 Rating | Score | Evaluator 2 Rating | Score |
|---|---|---|---|---|---|
| Data Processing | 1 | Rating: 5<br><br>Justification: The testbed's ability to do data processing was not explored in the article, but it is assumed that data could be exported for analysis. | 5 | Rating: 5<br><br>Justification: The nature of the Wu et al. (2021) work suggests that it is possible to export captured data. No visualization tools seem to exist. | 5 |
| System Architecture | 1 | Rating: 5<br><br>Justification: The system architecture is open-source and code is available. | 5 | Rating: 5<br><br>Justification: The system is open source. It was not possible to determine the quality of its architecture, but it should not present insurmountable challenges. There do not appear to be any SDKs or other tools to support system maintenance/modification. | 5 |

| Evaluation Criteria | Weight | Evaluator 1 Rating | Score | Evaluator 2 Rating | Score |
|---|---|---|---|---|---|
| Agents | 1 | Rating: 8<br><br>Justification: Three types of agents were studied using this testbed<br><br>- Agents worked in parallel efficiently and individually.<br>- Agents cooperated on the same subtasks when necessary.<br>- Agents stayed out of each other's way.<br><br>This demonstrates the ability to "plug in" agents with different abilities. | 8 | Rating: 6<br><br>Justification: Wu et al. (2021) developed agents and embedded them within the game. There was no mention of doing this via APIs. | 6 |
| Total | | | 62/150 | | 66/150 |

The unweighted scores for each evaluation dimension are shown in Figure 19.

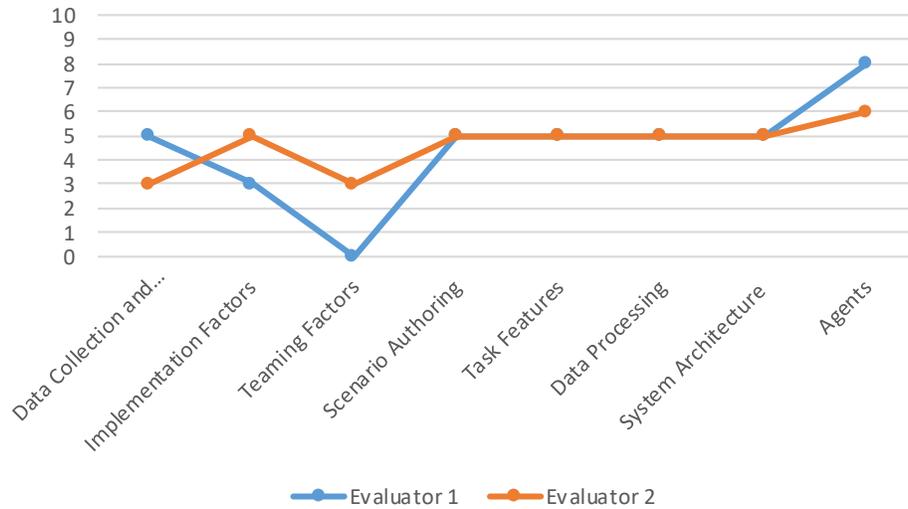

**Figure 19: Evaluation Dimension Scores for Overcooked!-based "Salad" Testbed**

Table 22:  *Neverwinter Nights* Evaluation

| Evaluation Criteria | Weight | Evaluator 1 Rating | Score | Evaluator 2 Rating | Score |
|---|---|---|---|---|---|
| Data Collection/ Performance Measures | 3 | Rating: 6<br><br>Justification: Built-in scripting API makes it possible to log many types of player actions in the database.  Bioware Inc. (the game developers) have also added time stamping and complete player-text-window logging capabilities.  Finally, teams increase performance scores by completing tasks for rewards (recovering caches), and managing costs and penalties (opening empty caches or setting off traps). | 18 | Rating: 8<br><br>Justification: Leung, Diller, & Ferguson (2005) noted, "the built-in scripting API made it possible to log many types of player actions." Therefore, there is reason to believe that data collection is relatively rich, but not unlimited. | 24 |

| Evaluation Criteria | Weight | Evaluator 1 Rating | Score | Evaluator 2 Rating | Score |
|---|---|---|---|---|---|
| Implementation Factors | 3 | Rating: 6<br><br>Justification: The goal with SABRE was to develop a flexible, customizable experimentation tool for behavior research, using a COTS program as the base.<br><br>*Neverwinter Nights* offers an immersive, interactive, virtual experience to reduce the work needed to create something entirely new. It has extensive API and a scripting language built in for researchers to modify it. It also regularly receives software updates, and is highly customizable.<br><br>Third-party software can also be integrated in the game to handle data (as was done in the SABRE testbed). | 18 | Rating: 6<br><br>Justification: SABRE employs the commercial game *Neverwinter Nights*. It is not open source, but various authoring tools and APIs exist to facilitate expansion. The environment is relatively complex. It seems likely that initially establishing a STE based on *Neverwinter Nights* would be associated with a relatively significant one-time level of effort. The on-going level of effort would be less extensive, but perhaps greater than that associated with the open-source environments. | 18 |

| **Evaluation Criteria** | **Weight** | **Evaluator 1 Rating** | **Score** | **Evaluator 2 Rating** | **Score** |
|---|---|---|---|---|---|
| Teaming Factors | 2 | Rating: 6<br><br>Justification: Researchers describe being able to assign different team members to roles (patrol leader, weapons specialist). Teammates are encouraged to distribute responsibilities, and are provided limited supplies that must be shared, as well as restricting the flow of information to mandate point-to-point communication.<br><br>The size of teams was not explicitly described, and team configuration appeared to focus on all-human teams. | 12 | Rating: 6<br><br>Justification: SABRE and *Neverwinter Nights* support teams with multiple players and various defined roles. Leung and her colleagues (2005) did not establish the minimum or maximum number of players or roles, but the presence of the scripting tools suggests a high-level of flexibility. SABRE/*Neverwinter Nights* include Non-Player Characters, suggesting that it should be possible to include AI-based agents within the environment. | 12 |

| Evaluation Criteria | Weight | Evaluator 1 Rating | Score | Evaluator 2 Rating | Score |
|---|---|---|---|---|---|
| Scenario Authoring | 2 | Rating: 7<br><br>Justification: SABRE (via *Neverwinter Nights*) presents a task with customizable goals, resources, events, and world maps in order to design experiments about particular behaviors/contexts. Researchers have developed scenarios for group planning and task execution, building in metrics for effectiveness and efficiency.<br><br>However, it is unclear how simple it is to design individual scenario events that can be linked to specific events. | 14 | Rating: 9<br><br>Justification: SABRE/*Neverwinter Nights* supports authoring. Game editing tools, the API, and community content provides a great deal of flexibility in this regard. | 18 |

| Task Features | 2 | Rating: 9 | 18 | Rating: 8 | 16 |
|---|---|---|---|---|---|
| | | Justification: Researchers acknowledged that they didn't need gameplay in the testbed to be "high fidelity" for military tasks. However, the task needed to represent situations that elicited behaviors with military relevance. SABRE's designers wanted a game that supported a wide variety of scenarios and contexts. The selection was narrowed to role-playing games to offer a larger scope for slower-paced deliberation, communication, and decision-making. Performance is team-based, with participants able to increase the team score by completing tasks (which have rewards) while managing costs and penalties. The researchers also wanted to be able to manipulate a game's scoring system so that the players demonstrate the weighing of costs and | | Justification: As a role-playing environment, SABRE and *Neverwinter Nights* should provide a good deal of flexibility with respect to the challenges presented to the participants and the settings in which those challenges take place (Leung, Diller, & Ferguson, 2005; Gorniak and Roy, 2005; Carbonaro et al., 2005). It took approximately one hour for players in the Leung, Diller, & Ferguson (2004) to become adequately familiar with the environment. They did not state how long a given session lasted. The scripting language should provide a great deal of control over scenario difficulty, information processing complexity, and related factors (*e.g.*, signal-to-noise ratio, data reliability, uncertainty). Similarly, the flexibility presented by SABRE/*Neverwinter Nights* suggests that challenges can be developed that require differing levels of interdependence and communication among teammates. However, employing this flexibility requires the use of the scripting environment, thus increasing development timelines and costs. In the version of *Neverwinter Nights* used within SABRE, players could only communicate using chat | |

| Evaluation Criteria | Weight | Evaluator 1 Rating | Score | Evaluator 2 Rating | Score |
|---|---|---|---|---|---|
| | | benefits in game strategy. Overall, many of the desired manipulations in the evaluation criteria appear possible within SABRE, through the extensive scenario authoring capabilities. | | utterances. We suspect, but have not confirmed, that voice-based communication is supported in more recent releases. | |
| Data Processing | 1 | Rating: 5 Justification: The game generates log files on the client side, which requires the testbed to be able to merge the information into a central database. The testbed cannot collect complete data from just the server. Information about data visualization was not presented in the testbed description. | 5 | Rating: 5 Justification: The SABRE team created an analysis toolkit for data management and analysis. This toolkit is not available to us and we would have to recreate it. | 5 |

| Evaluation Criteria | Weight | Evaluator 1 Rating | Score | Evaluator 2 Rating | Score |
|---|---|---|---|---|---|
| System Architecture | 1 | Rating: 7<br><br>Justification: The game uses a client-server architecture where the server is responsible for maintaining the game state and the client supports the user interface for viewing and interacting with the game world. | 7 | Rating: 7<br><br>Justification: SABRE is based on *Neverwinter Nights*, a closed-source commercial game that includes game editing tools and APIs. *Neverwinter Nights* is organized as a client-server system and the SABRE team added both client- and server-side components to create the testbed. The architecture of the base game is not relevant, since we cannot change it. However, there are SDKs and/or other tools to support system maintenance/modification. | 7 |

| Evaluation Criteria | Weight | Evaluator 1 Rating | Score | Evaluator 2 Rating | Score |
|---|---|---|---|---|---|
| Agents | 1 | Rating: 2<br><br>Justification: SABRE was envisioned as an environment to demonstrate and test synthetic entities, however those entities were not described as teammates to humans.<br><br>It is unclear whether they could be programmed to function as teammates.<br><br>The synthetic entities are engaged with using a communication menu. This places a limit on communication within the game between humans and agents. | 2 | Rating: 6<br><br>Justification: *Neverwinter Nights* includes non-player characters and the SABRE team modified the behavior of the NPCs. This suggests that it should be possible to interface the game environment with arbitrary agent technology. However, Leung, Diller, & Ferguson (2005) did not do so when SABRE was developed. | 6 |
| Total | | | 94/150 | | 106/150 |

The unweighted scores for each evaluation dimension are shown in Figure 20.

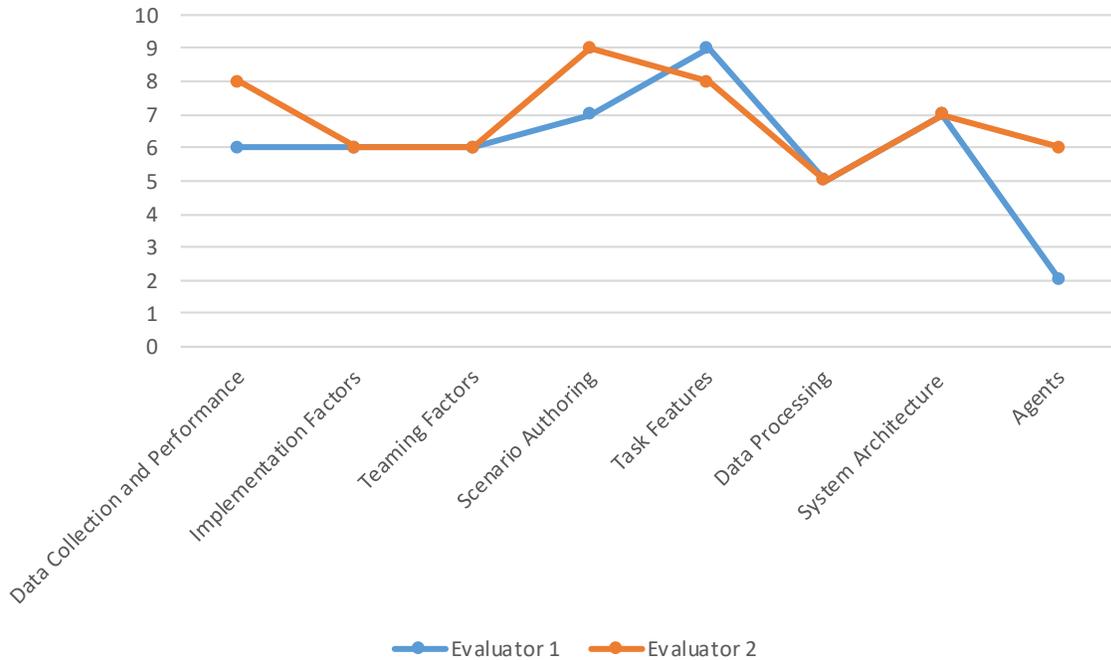

**Figure 20: Evaluation Dimension Scores for *Neverwinter Nights***

Table 23: ASIST Saturn+ Testbed Evaluation

| Evaluation Criteria | Weight | Evaluator 1 Rating | Score | Evaluator 2 Rating | Score |
|---|---|---|---|---|---|
| Data Collection/ Performance Measures | 3 | Rating: 8<br><br>Justification: The ASIST testbed collected a wide array of outcome and process variables, as well as conducting complex NLP for agent communication within the testbed. While many measures appeared to be available, it was not obvious how simple it would be to program fully unique measures, but it is likely possible with some work. | 24 | Rating: 10<br><br>Justification: The Analytic Components described in the research protocol seem to have unlimited flexibility. The instrumentation seems impressive. | 30 |

| Evaluation Criteria | Weight | Evaluator 1 Rating | Score | Evaluator 2 Rating | Score |
|---|---|---|---|---|---|
| Implementation Factors | 3 | Rating: 6<br><br>Justification: This testbed has a lot to offer on the outset, meaning it would likely be very easy to integrate into the testbed, however it is not clear how simple it would be to create new missions sets. There is likely some cost associated with that, given the lack of available authoring tools. | 18 | Rating: 7<br><br>Justification: The system seems to have a well-architected modular architecture. However, it does not have robust authoring tools or SDKs. | 21 |
| Teaming Factors | 2 | Rating: 4<br><br>Justification: Team size is on the smaller end but there appear to be clearly defined roles. Team communication is measured and analyzed to an impressive degree. The downside is that only humans appear to be able to fill the roles outlined in the description. | 8 | Rating: 4<br><br>Justification: The team size is limited to three and it appears that all three must be human. However, the availability of the coaching agents suggest that it may be possible to add virtual teammates to the system. | 8 |

| Evaluation Criteria | Weight | Evaluator 1 Rating | Score | Evaluator 2 Rating | Score |
|---|---|---|---|---|---|
| Scenario Authoring | 2 | Rating: 3<br><br>Justification: There were no scenario authoring tools available. Instead one is able to determine when missions experience perturbations. Modifying scenarios beyond that likely requires significant work. | 6 | Rating: 2<br><br>Justification: ASIST Saturn+ does not include authoring tools. However, there are factors within the two defined scenarios that suggest that such tools are possible. It is not clear how difficult creation of those tools would be. | 4 |
| Task Features | 2 | Rating: 7<br><br>Justification: The task not only has high fidelity with actual search and rescue environments, it appears to present many complex coordination and decision-making challenges to successfully manage effective teaming performance to optimize team members' performance. The environment seems to present a moderate ability to modulate difficulty. | 14 | Rating: 6<br><br>Justification: It appears that ASIST Saturn+ scenarios include a number of features that developers could use control scenario difficulty. There is at least a moderate level of information processing and decision-making required by the task and (especially) by the perturbations. | 12 |

| Evaluation Criteria | Weight | Evaluator 1 Rating | Score | Evaluator 2 Rating | Score |
|---|---|---|---|---|---|
| Data Processing | 1 | Rating: 7<br><br>Justification: The testbed features analytic components that support performance assessment. It is assumed that data can be exported using conventional formats, but none are discussed directly. No data visualization appears to be available. | 7 | Rating: 8<br><br>Justification: ASIST Saturn+ includes Analytic Components that can be designed to provide robust analysis of user performance. Data visualization does not appear to be supported. We assume that data can be exported, but the nature of that process is unclear. | 8 |
| System Architecture | 1 | Rating: 5<br><br>Justification: The testbed is based on COTS game *Minecraft*, meaning the software architecture is generally fixed, monolithic, and not amenable to significant changes. The research team developed additional software that was added to the game that appears to be somewhat modular. | 5 | Rating: 7<br><br>Justification: ASIST Saturn+ is based on *Minecraft*, a closed-source commercial game. The ASIST Saturn+ team added components to the Minecraft foundation to create the testbed. The architecture of the base game is not relevant, since we cannot change it. However, the support features appear to be modular and well-architected. | 7 |

| Evaluation Criteria | Weight | Evaluator 1 Rating | Score | Evaluator 2 Rating | Score |
|---|---|---|---|---|---|
| Agents | 1 | Rating: 5<br><br>Justification: Agents can likely be incorporated into the system, but doing so will take a significant software development effort to achieve. | 5 | Rating: 5<br><br>Justification: ASIST Saturn+ currently only supports agents in the advisor role, not the player role. However, since there is support for agents of some type, we believe that it should be possible to integrate agents as virtual teammates. | 5 |
| Total | | | 87/150 | | 95/150 |

The unweighted scores for each evaluation dimension are shown in Figure 21.

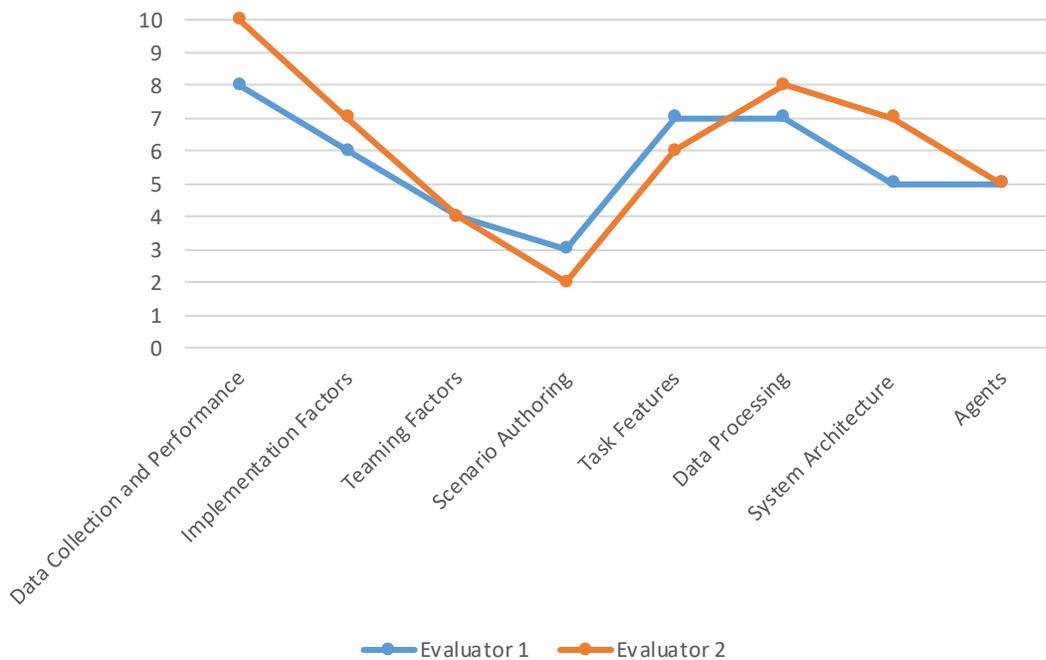

Figure 21: Evaluation Dimension Scores for ASIST Saturn+ Testbed

Table 24: ASIST Dragon Testbed Evaluation

| Evaluation Criteria | Weight | Evaluator 1 Rating | Score | Evaluator 2 Rating | Score |
|---|---|---|---|---|---|
| Data Collection/ Performance Measures | 3 | Rating: 9<br><br>Justification: Many data points are able to be collected in this testbed. In addition, intake surveys and post-trial surveys are available, and communication is recorded for analysis. | 27 | Rating: 10<br><br>Justification: The Analytic Components described in the research protocol seem to have unlimited flexibility. The instrumentation seems impressive. | 30 |
| Implementation Factors | 3 | Rating: 8<br><br>Justification: The ASIST Dragon testbed would likely be reasonably simple to integrate. There are also preliminary authoring options associated with this testbed that give it more flexibility. There are no SDKs or specific authoring tools. | 24 | Rating: 7<br><br>Justification: The system seems to have a well-architected modular structure. However, it does not have robust authoring tools or SDKs. | 21 |
| Teaming Factors | 2 | Rating: 3<br><br>Justification: This testbed supports all-human three-member teams without role differentiation.<br><br>Communication is recorded (through text chat, and use of communication beacons). | 6 | Rating: 3<br><br>Justification: The team size is limited to three. The roles are not differentiated. It appears that all three must be human. However, the availability of the coaching agents suggest that it may be possible to add virtual teammates to the system | 6 |

| Evaluation Criteria | Weight | Evaluator 1 Rating | Score | Evaluator 2 Rating | Score |
|---|---|---|---|---|---|
| Scenario Authoring | 2 | Rating: 6<br><br>Justification: This testbed allows for certain changes to the task environment (bomb layouts and different region selection).<br><br>Specifically, modifications of the task and environment are designed to fit different domain research purposes including training human teams, developing AI teammates. However, it is unclear how expressive the task can be made to be. | 12 | Rating: 6<br><br>Justification: Dragon does not include authoring tools. However, there are factors within the defined scenarios that suggest that such tools are possible. It is not clear how difficult creation of those tools would be. | 12 |
| Task Features | 2 | Rating: 7<br><br>Justification:<br>The tasking in the testbed is designed to study team dynamics, and includes cognitive team tasking such as planning and negotiating tasks, perturbations that can be introduced during normal tasking, and has optional interdependent team tasks that can be layered into tasking. The task environment has high fidelity to a typical mission environment and the tasking presents reasonable information processing challenges to teams. | 14 | Rating: 6<br><br>Justification: It appears that Dragon scenarios include a number of features that developers could use control scenario difficulty. There is at least a moderate level of information processing and decision-making required by the task and the perturbations. The use of a planning phase is an interesting aspect of the environment. | 12 |

| Evaluation Criteria | Weight | Evaluator 1 Rating | Score | Evaluator 2 Rating | Score |
|---|---|---|---|---|---|
| Data Processing | 1 | Rating: 7<br><br>Justification: The Dragon testbed has robust analytic components to support performance assessment. It is assumed that data can be exported using conventional formats (none are mentioned directly). No data visualization appears to be available | 7 | Rating: 8<br><br>Justification: Dragon includes Analytic Components that can be designed to provide robust analysis of user performance. Data visualization does not appear to be supported. We assume that data can be exported, but the nature of that process is unclear. | 8 |

| Evaluation Criteria | Weight | Evaluator 1 Rating | Score | Evaluator 2 Rating | Score |
|---|---|---|---|---|---|
| System Architecture | 1 | Rating: 6<br><br>Justification: The testbed is based on COTS game *Minecraft*, meaning the software architecture is not amenable to changes. The research team developed additional software that was added to the game that appears to be somewhat modular. Further, servers are available 24/7 without need for administrative support for data collection. | 6 | Rating: 7<br><br>Justification: Dragon is based on *Minecraft*, a closed-source commercial game. Earlier ASIST testbeds attempted to use an open source version of *Minecraft*, but technical difficulties were encountered and a similar version of the commercial software was adopted. The Dragon team added components to the *Minecraft* foundation create the testbed. Those elements are available under open-source licenses. The architecture of the base game is not relevant, since we cannot change it. However, the support features appear to be modular and well-architected. | 7 |

| Evaluation Criteria | Weight | Evaluator 1 Rating | Score | Evaluator 2 Rating | Score |
|---|---|---|---|---|---|
| Agents | 1 | Rating: 5<br><br>Justification: The testbed involves real time ASI advisor interventions (driven by social science-influenced analytic components), rather than AI teammates. However, one goal of the task environment is to be able to support developing AI teammates. This support means it is likely that agents can be integrated as teammates even if they are not currently. | 5 | Rating: 5<br><br>Justification: Dragon currently only supports agents in the advisor role, not the player role. However, since there is support for agents of some type, we believe that it should be possible to integrate agents as virtual teammates. | 5 |
| Total | | | 101/150 | | 101/150 |

The unweighted scores for each evaluation dimension are shown in Figure 22.

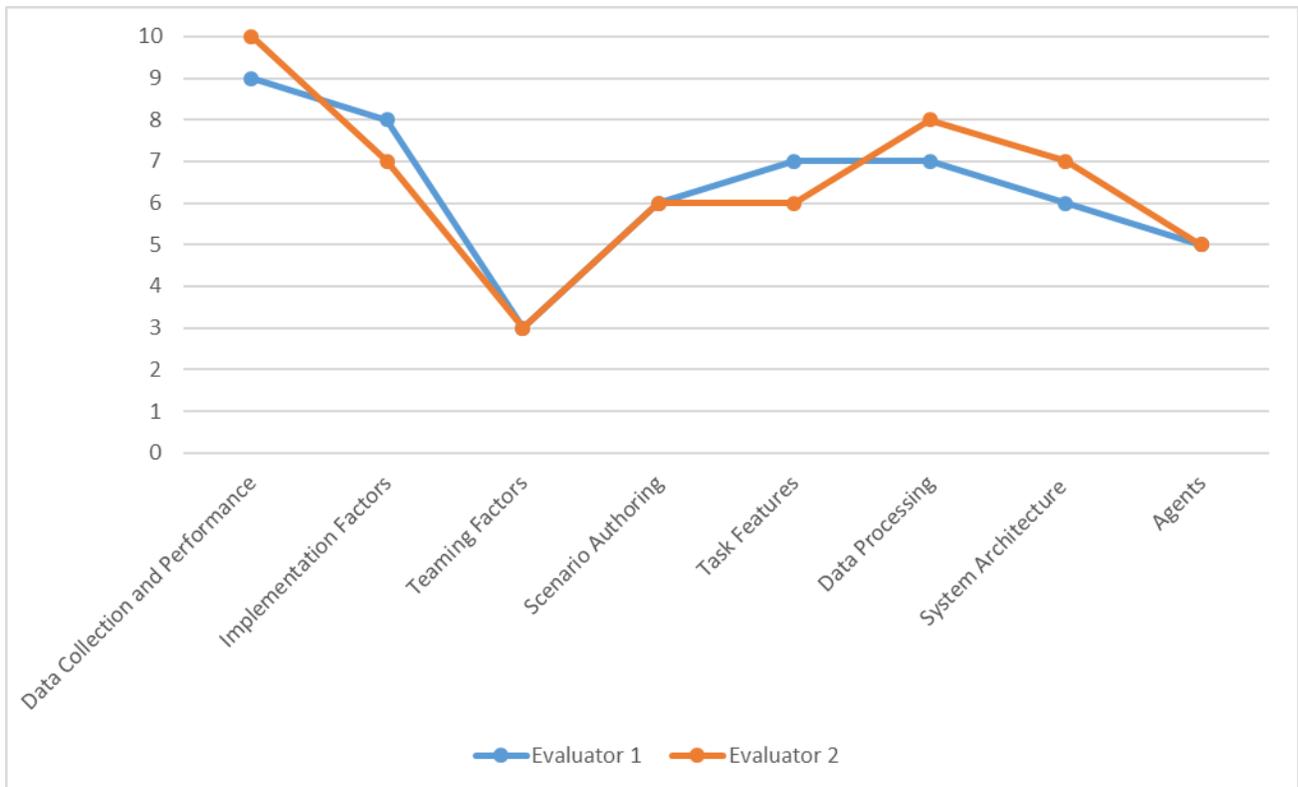

**Figure 22: Evaluation Dimension Scores for ASIST Dragon Testbed**